\documentclass[12pt]{article}
\usepackage{epsfig}
\usepackage{amsmath,amssymb,amsthm,amscd}
\newcommand{\twoVgraph}{\raisebox{0pt}{
                 \begin{picture}(18,18)(-9,-5)
                 \put(0,0){\circle{16}} \put(-8,0){\line(1,0){16}}
                 \end{picture}}}

\newcommand{\IP}{\relax{\rm I\kern-.18em P}}
\newcommand{\IR}{\relax{\rm I\kern-.18em R}}

\newcommand{\tr}{{\rm Tr}}
\newcommand{\CQ}{{\cal Q}}
\newcommand{\zb}{\bar z}
\newcommand{\be}{\begin{equation}}
\newcommand{\ee}{\end{equation}}
\newcommand{\ben}{\begin{eqnarray}\displaystyle}
\newcommand{\een}{\end{eqnarray}}

\newcommand{\sectiono}[1]{\section{#1}\setcounter{equation}{0}}

\newcommand{\figref}[1]{Fig.~\protect\ref{#1}}

\begin{document}
{}~
\hfill\vbox{\hbox{hep-th/0410165}
\hbox{CERN-PH-TH/2004-199}
}\break

\vskip .6cm

\centerline{\Large \bf
Les Houches lectures on matrix models}\vskip .1cm
\centerline{\Large \bf and topological strings}

\medskip

\vspace*{4.0ex}

\centerline{\large \rm
Marcos Mari\~no\footnote{Also at Departamento de
Matem\'atica, IST, Lisboa, Portugal.}}

\vspace*{4.0ex}

\centerline{ Department of Physics, CERN}
\centerline{  Geneva 23, CH-1211 Switzerland}
\vspace*{2.0ex}
\centerline{marcos@mail.cern.ch}

\vspace*{5.0ex}

\centerline{\bf Abstract} \bigskip

In these lecture notes for the Les Houches School on 
Applications of Random Matrices in Physics we give an introduction to the connections between matrix models 
and topological strings. We first review some basic results of matrix model 
technology and then we focus on type B topological strings. We present the main 
results of Dijkgraaf and Vafa 
describing the spacetime string dynamics on certain Calabi-Yau backgrounds in terms 
of matrix models, and we emphasize the connection to geometric transitions and 
to large $N$ gauge/string duality. We also use matrix model technology 
to analyze large $N$ Chern-Simons theory and the Gopakumar-Vafa transition.

\vfill \eject

\baselineskip=16pt

\tableofcontents

\sectiono{Introduction}

Topological string theory was introduced by Witten in \cite{topmodel,topphase} as a simplified model 
of string theory which captures topological information of the target space, and it has been intensively 
studied since then. There are three important lessons that have been learned in the last few years about 
topological strings:

1) Topological string amplitudes are deeply related to 
physical amplitudes of type II string theory. 

2) The spacetime description 
of open topological strings in terms of string field theory reduces in some 
cases to very simple gauge theories.

3) There is an open/closed topological string duality which relates open and closed 
string backgrounds in a precise way 

In these lectures we will analyze a particular class of topological string theories where 
the gauge theory description in (2) above reduces in fact to a matrix model. This was 
found by Dijkgraaf and Vafa in a series of beautiful papers \cite{dv,dvtwo,dvthree}, where they 
also showed that, thanks to the connection to physical strings mentioned in (1),
the computation of nonperturbative superpotentials in a wide class of ${\cal N}=1$ gauge theories 
reduces to perturbative computations in a matrix model. This aspect of the work of Dijkgraaf and Vafa was very 
much explored and exploited, and rederived in the context of 
supersymmetric gauge theories without using the connection to topological strings. 
In these lectures 
we will focus on the contrary on (2) and (3), emphasizing the string field theory construction and the open/closed 
string duality. The applications of the results of Dijkgraaf and Vafa to supersymmetric gauge theories have been 
developed in many papers and reviewed for example in \cite{afh}, and we will not cover them here. 
Before presenting the relation between matrix models and topological strings, 
it is worthwhile to give a detailed conceptual discussion of the general ideas behind (2) and (3) and their 
connections to large $N$ dualities.

In closed string theory we study maps from a Riemann surface $\Sigma_g$ to a target manifold $X$, and 
the quantities we want to compute are the free energies at genus $g$, denoted by $F_g(t_i)$. Here, the $t_i$ 
are geometric data of the target space $X$, and the free energies are computed as correlation functions 
of a two-dimensional conformal field theory coupled to gravity. In topological string theory there are two 
different models, the A and the B model, the target space is a Calabi-Yau manifold (although this condition 
can be relaxed in the A model), and the parameters $t_i$ are K\"ahler and complex parameters, respectively. 
The free energies are assembled together into a generating functional
\be
\label{freeclosed}
F(g_s,t_i)=\sum_{g=0}^{\infty} g_s^{2g-2}F_{g}(t_i),
\ee
where $g_s$ is the string coupling constant. 

In open string theory we study maps from an open Riemann surface $\Sigma_{g,h}$ to a target $X$, 
and we have to provide boundary conditions as well. For example, we can impose Dirichlet conditions 
by using a submanifold $S$ of $X$ where the open strings have to end. In addition, we can use Chan-Paton factors 
to introduce a $U(N)$ gauge symmetry. The open string amplitudes are now $F_{g,h}$, and in the cases 
that will be studied in these lectures the generating functional will have the form 
\be
\label{freeopen}
F(g_s, N)=\sum_{g=0}^{\infty}\sum_{h=1}^{\infty} F_{g,h}g_s^{2g-2+h} N^h.
\ee
Physically, the introduction of Chan-Paton factors and boundary conditions through a submanifold $S$  
of $X$ means that we are wrapping $N$ (topological) D-branes around $S$. A slightly more general situation 
arises when there are $n$ submanifolds $S_1, \cdots, S_n$ where the strings can end. In this case, the open 
string amplitude is of the form $F_{g,h_1, \cdots, h_n}$ and the total free energy 
is now given by
\be
F(g_s, N_i)=\sum_{g=0}^{\infty}\sum_{h_1, \cdots, h_n=1}^{\infty} F_{g,h_1, \cdots, h_n} g_s^{2g-2+h} N_1^{h_1} 
\cdots N_n^{h_n},     
\label{fmulti}
\ee
where $h=\sum_{i=1}^n h_i$. In the case of open strings one can in some situations use string field theory to describe the spacetime dynamics. 
The open string field theory of Witten \cite{sft}, which was originally constructed for the open bosonic 
string theory, can also be applied to topological string theory, and on some particular Calabi-Yau backgrounds 
the full string field theory of the topological string reduces to a simple 
$U(N)$ gauge theory, where $g_s$ plays the role of the gauge coupling constant 
and $N$ is the rank of the gauge group. In particular, the string field reduces in this case to a finite number of 
gauge fields. As a consequence of this, the open string theory amplitude $F_{g,h}$ can 
be computed from the gauge theory by doing perturbation theory in the double line notation of 't Hooft 
\cite{thooft}. More precisely, $F_{g,h}$ is the contribution of the fatgraphs of genus $g$ and $h$ holes. 
The idea that fatgraphs of a $U(N)$ gauge theory correspond to open string amplitudes is an old one, and 
it is very satisfying to find a concrete realization of this idea in the context of a string field 
theory description of topological strings, albeit for rather simple gauge theories.  

The surprising fact that the full string field theory is described by a simple gauge theory is typical 
of topological string theory, and does not hold for conventional string models. 
There are two examples where this description has been worked out:

1) The A model on a Calabi-Yau of the form $X=T^*M$, where $M$ is a three-manifold, and there are $N$ 
topological D-branes wrapping $M$. In this case, the gauge theory is Chern-Simons theory on $M$ \cite{csts}.

2) The B model on a Calabi-Yau manifold $X$ which is the small resolution of a singularity characterized 
by the hyperelliptic curve $y^2=(W'(x))^2$. If $W'(x)$ has degree $n$, the small resolution produces 
$n$ two-spheres, and one can wrap $N_i$ topological D-branes around each two-sphere, with $i=1, \cdots, n$. 
In this case Dijkgraaf and Vafa showed that the gauge theory is a multicut matrix model with potential 
$W(x)$ \cite{dv}.

In both examples, the open string amplitudes $F_{g,h}$ 
are just numbers computed by the fatgraphs of the 
corresponding gauge theories.  

The fatgraph expansion of a $U(N)$ gauge theory can be resummed formally by introducing the so called 
't Hooft parameter $t=g_s N$. For example, in the case of the free energy, we can rewrite (\ref{freeopen}) in 
the form (\ref{freeclosed}) by defining
\be
\label{thooftre}
F_g(t)=\sum_{h=1}^{\infty} F_{g,h} t^h.
\ee
In other words, starting from an {\it open} string theory expansion we can obtain 
a {\it closed} string theory expansion by resumming the hole expansion as indicated in (\ref{thooftre}). 
This idea was proposed by 't Hooft \cite{thooft} and gives a closed string theory interpretation 
of a gauge theory. 

What is the interpretation of the above resummation for the 
gauge theories that describe the spacetime dynamics of topological open string theories? 
As was explained in \cite{gv} 
(for the A model example above) and in \cite{civ} (for the B model example), there is a 
{\it geometric} or {\it large $N$} transition that relates the open string Calabi-Yau background 
$X$ underlying the gauge theory to a closed string Calabi-Yau background $X'$. The geometric transition typically 
relates two different ways of smoothing out  a singular geometry (the ``resolved'' geometry and the 
``deformed'' geometry). Moreover, the ``master field'' that describes the large $N$ limit \cite{master} 
turns out to encode the target space geometry of the closed string background, and the 't Hooft parameter becomes 
a geometric parameter of the resulting closed geometry. 
\begin{figure}[!ht]
\leavevmode
\begin{center}
\epsfysize=6.5cm
\epsfbox{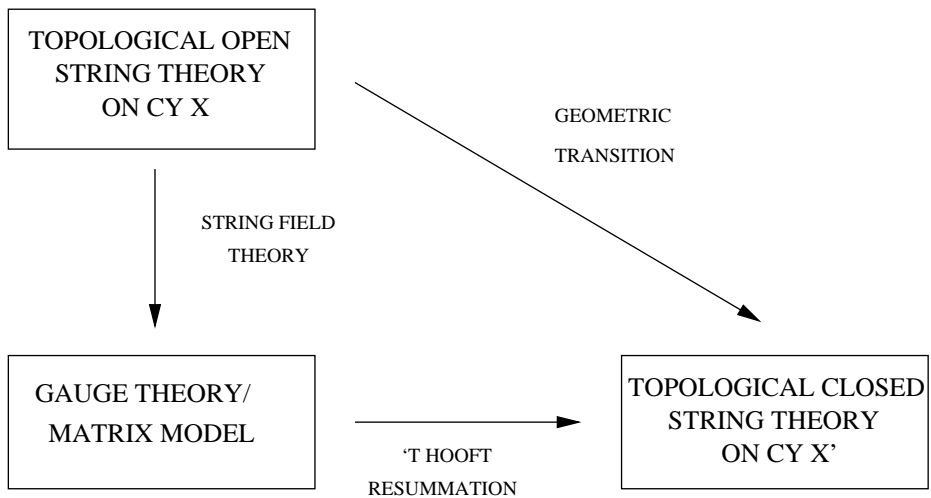}
\end{center}
\caption{This diagram summarizes the different relations between closed topological strings, open 
topological strings, and gauge theories.}
\label{concept}
\end{figure} 
The idea that an open string background with D-branes is equivalent to a different, 
geometric closed string background (therefore with no D-branes) appeared originally in the 
AdS/CFT correspondence \cite{malda}. In this correspondence, type IIB theory in flat space in the presence of D-branes is
conjectured to be equivalent to type IIB theory in ${\rm AdS}_5 \times {\bf
S}^5$ with no D-branes, and where the radius of the ${\bf S}^5$ is related
to the 't Hooft parameter. The reason this holds is that, at large $N$, the
presence of the D-branes can be traded for a deformation of the background
geometry. In other words, we can make the branes disappear if we change the
background geometry at the same time. Therefore, as emphasized by Gopakumar and Vafa in \cite{gv},
large $N$ dualities relating open and closed strings should be associated
to transitions in the geometry. The logical structure of all the connections we have sketched 
is depicted in \figref{concept}. 
        
In these lectures we will mostly focus on the B topological string, the Dijkgraaf-Vafa scenario, and 
the geometric transition of \cite{civ}. For a detailed review of a similar story for the 
A string, we refer the reader to \cite{mmreview}. The organization of these lectures is as follows. In section 2 we review some basic ingredients 
of matrix models, including saddle-point techniques and orthogonal polynomials. In section 3 we explain 
in detail the connection between matrix models and topological strings due to 
Dijkgraaf and Vafa. We first review the topological B model and its string field theory description, and we show that in the 
Calabi-Yau background associated to the resolution of a polynomial singularity, the string field theory 
reduces to a matrix model. We develop some further matrix model technology to understand all the 
details of this description, and we make the connection with geometric transitions. In section 4 we briefly 
consider the geometric transition of Gopakumar and Vafa \cite{gv} from the point of view of the matrix model 
description of Chern-Simons theory. This allows us to use matrix model technology to derive 
some of the results of \cite{gv}.

\sectiono{Matrix models}

In this section we develop some aspects and techniques of matrix models 
which will be needed in the following. There are excellent reviews of this material, 
such as for example \cite{df,dfgz}.   

\subsection{Basics of matrix models}

Matrix models are the simplest examples of quantum gauge theories, 
namely, they are quantum gauge theories in zero dimensions. The basic field is
a Hermitian $N \times N$ matrix $M$.  
We will consider an action for $M$ of the
form:
\be
\label{genmm}
{1\over g_s}W(M)={1 \over 2g_s}{\rm Tr}\, M^2+ {1\over g_s} \sum_{p\ge 3} {g_p\over p} {\rm Tr}\, M^p.
\ee
where $g_s$ and $g_p$ are coupling constants. This action has the obvious gauge symmetry
\be
M \rightarrow U M U^{\dagger},\ee
where $U$ is a $U(N)$ matrix. The partition function of the theory is given by
\be
\label{matrix}
Z={1 \over {\rm vol}(U(N))}\int dM \, e^{-{1\over g_s}W(M)}
\ee
where the factor ${\rm vol}(U(N))$ is the usual volume factor of
the gauge group that arises after
fixing the gauge. In other words, we are considering here a {\it gauged} matrix model.
The measure in the ``path integral'' is the Haar measure
\be
\label{haarmeas}
dM = 2^{N(N-1)\over 2}\prod_{i=1}^N dM_{ii} \prod_{1\le i<j\le N} d{\rm Re}\, M_{ij} d {\rm Im}\, M_{ij}.
\ee
The numerical factor in (\ref{haarmeas}) is introduced to obtain a convenient normalization.

A particularly simple example is the {\it Gaussian matrix model}, defined by the partition function
\be
\label{gmm}
Z_G=
{1  \over {\rm vol}(U(N))}\int dM \, e^{-{1 \over 2g_s} {\rm Tr} \, M^2}.
\ee
We will denote
by
\be
\label{vevg}
\langle f(M) \rangle_G ={\int dM \,f(M)\,  e^{-{\rm Tr}\,  M^2/2g_s} \over \int dM \,  e^{-{\rm Tr}\,  M^2/2g_s}}
\ee
the normalized vevs of a gauge-invariant functional $f(M)$ in the Gaussian matrix model. 
This model is of course exactly solvable, and the vevs (\ref{vevg}) can be computed systematically as 
follows. Any gauge-invariant function $f(M)$ can be written as a linear combination of traces of $M$ in arbitrary representations $R$  
of $U(N)$. If we represent $R$ by a Young tableau with rows of lengths $\lambda_i$, with $\lambda_1 \ge \lambda_2 \ge \cdots$, and 
with $\ell(R)$ boxes in total, we define the set of $\ell(R)$ integers $f_i$ as follows
\begin{equation}
\label{fis}
f_i=\lambda_i +\ell(R) -i,\,\,\,\,\,\, i=1, \cdots, \ell(R).
\end{equation}
Following \cite{idf}, we will say that the Young tableau associated to
$R$ is even if the number of odd $f_i$'s is
the same as the number of even $f_i$'s. Otherwise, we will say that it is
odd. If $R$ is even, one has the following 
result \cite{iz,idf}:
\begin{equation}
\label{averun}
\langle {\rm Tr}_R M\rangle_G = c(R)\, {\rm dim}\, R,
\end{equation}
where 
\be
c(R)=(-1)^{A(A-1)\over 2}
{\prod_{f \, {\rm odd}} f!! \prod_{f' \, {\rm even}} f'!!
\over \prod_{f \, {\rm odd}, f'\, {\rm even}} (f-f')} 
\ee
and $A=\ell(R)/2$ (notice that $\ell(R)$ has to be even in order to have
a non-vanishing result). Here ${\rm dim}\, R$ is the dimension of the
irreducible representation of $SU(N)$ associated to $R$,
and can be computed for example by using the hook formula. On the other hand, if $R$ is odd, 
the above vev vanishes. 

The partition function $Z$ of more general matrix models with action (\ref{genmm}) 
can be evaluated by doing perturbation theory around
the Gaussian point: one expands the exponential of
$\sum_{p\ge 3} (g_p/g_s) {\rm Tr} M^p/p$ in (\ref{matrix}), and computes the partition function
as a power series in the coupling constants $g_p$. The evaluation of each term of the 
series involves the computation of vevs like (\ref{vevg}). Of course, this computation
can be interpreted in terms of Feynman diagrams, and as usual the perturbative expansion of the free energy
$$
F=\log \, Z$$
will only involve connected vacuum bubbles. 

\begin{figure}[!ht]
\leavevmode
\begin{center}
\epsfysize=1cm
\epsfbox{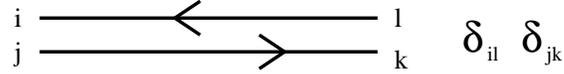}
\end{center}
\caption{The propagator in the double line notation.}
\label{propa}
\end{figure}

\begin{figure}[!ht]
\leavevmode
\begin{center}
\epsfysize=4cm
\epsfbox{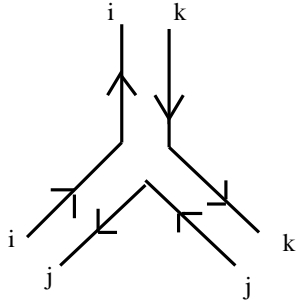}
\end{center}
\caption{The cubic vertex in the double line notation.}
\label{cubicvertex}
\end{figure}

\begin{figure}[!ht]
\leavevmode
\begin{center}
\epsfysize=4cm
\epsfbox{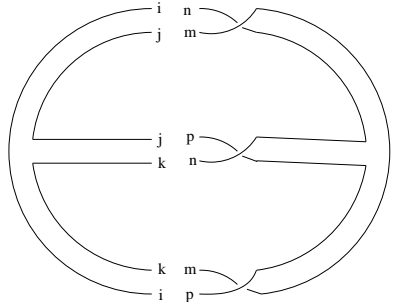}
\end{center}
\caption{Contracting two cubic vertices in the double line notation: the $N$ contribution.}
\label{nsolo}
\end{figure}

\begin{figure}[!ht]
\leavevmode
\begin{center}
\epsfysize=3cm
\epsfbox{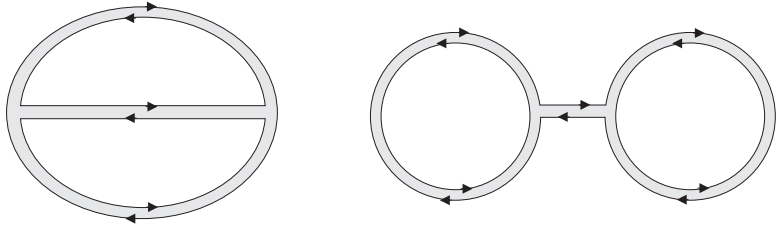}
\end{center}
\caption{Two planar diagrams in the cubic matrix model.}
\label{leading}
\end{figure}

Since we are dealing with a quantum theory of a field in the adjoint representation we can 
reexpress the perturbative expansion of $F$ in terms of fatgraphs, by using the double line notation due to 
't Hooft \cite{thooft}. The purpose of the fatgraph expansion is the following: in $U(N)$ gauge theories 
there is, in addition to the coupling constants appearing in the model (like for example $g_s, g_p$ in (\ref{genmm})), 
a hidden variable, namely $N$, the rank of the gauge group. The $N$ dependence in 
the perturbative expansion comes from the group factors associated to Feynman diagrams,
but in general a single Feynman diagram gives rise to a polynomial in $N$ involving
different powers of $N$. Therefore, the standard Feynman diagrams, which are good in order
to keep track of powers of the coupling constants, are not good in order to keep track of powers
of $N$. If we want to keep track of the $N$ dependence we have to ``split'' each diagram into different pieces which
correspond to a definite power of $N$. To do that, one
writes the Feynman diagrams of the theory as ``fatgraphs'' or double line graphs, as first indicated by 't Hooft \cite{thooft}.
\begin{figure}[!ht]
\leavevmode
\begin{center}
\epsfysize=1cm
\epsfbox{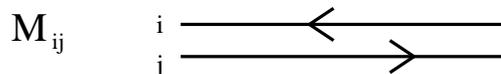}
\end{center}
\caption{The index structure of the field $M_{ij}$ in the adjoint representation of
$U(N)$ is represented through a double line.}
\label{fat}
\end{figure} 
Let us explain this in some detail, taking the example of the matrix model with a cubic 
potential ({\it i.e.} $g_p=0$ in (\ref{genmm}) for $p>3$). The fundamental field
$M_{ij}$ is in the adjoint representation. Since the adjoint representation of $U(N)$ is the tensor
product of the fundamental $N$ and the antifundamental ${\overline N}$, we can look at $i$ (resp. $j$)
as an index of the fundamental (resp. antifundamental) representation. We will represent
this double-index structure by a double line notation as shown in \figref{fat}. The only thing
we have to do now is to rewrite the Feynman rules of the theory by 
taking into account this double-line notation. For example, the kinetic term of the theory
is of the form
\be
{1 \over g_s} {\rm Tr}\, M^2 ={1 \over g_s} \sum_{i,j} M_{ij} M_{ji}.
\ee
This means that the propagator of the theory is
\be
\langle M_{ij} M_{kl} \rangle = g_s \delta_{il} \delta_{jk}
\ee
and can be represented in the double line notation as in \figref{propa}. 
Next, we consider the vertices of the theory. For example, the trivalent vertex given by
\be
{g_3\over g_s} {\rm Tr}\, M^3 ={g_3 \over g_s}\sum_{i,j,k} M_{ij}\, M_{jk} \, M_{ki}
\ee
can be represented in the double line notation as in \figref{cubicvertex}. A vertex of order $p$ 
can be represented in a similar way by drawing $p$ double lines joined together.
Once we have rewritten the Feynman rules in the double-line notation, we can construct the
corresponding graphs, which look like ribbons and are called ribbon graphs or fatgraphs.
It is clear that in general a usual Feynman diagram can give rise to many different fatgraphs.
Consider for example the one-loop vacuum diagram $\twoVgraph$, which
comes from contracting two cubic vertices. 
\begin{figure}[!ht]
\leavevmode
\begin{center}
\epsfysize=4cm
\epsfbox{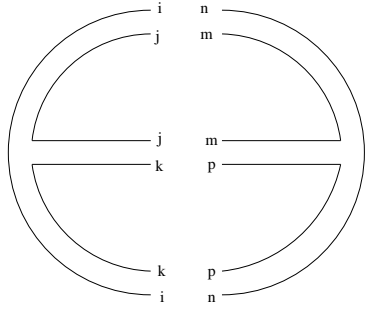}
\end{center}
\caption{Contracting two cubic vertices in the double line notation: the $N^3$ contribution.}
\label{ncube}
\end{figure}
In the double line notation the contraction
can be done in two different ways. The first one is illustrated in \figref{ncube}
and gives a factor
\be
\sum_{ijkmnp}\langle M_{ij} M_{mn} \rangle \langle M_{jk} M_{pm} \rangle
\langle M_{ki} M_{np} \rangle = g_s^3 N^3.\ee
The second one is shown in \figref{nsolo} and gives a factor
\be
\sum_{ijkmnp}\langle M_{ij} M_{mn} \rangle \langle M_{jk} M_{np} \rangle
\langle M_{ki} M_{pm} \rangle =g_s^3 N.\ee
In this way we have split the original diagram into two different fatgraphs with a well-defined
power of $N$ associated to them. The number of factors of $N$ is simply equal to the number
of closed loops in the graph: there are three closed lines in the fatgraph resulting from the
contractions in \figref{ncube} (see the first graph in \figref{leading}), while there is
only one in the diagram resulting from \figref{nsolo}.
In general, fatgraphs turn out to be
characterized topologically by the number of propagators or edges $E$, 
the number of vertices with $p$ legs $V_p$, and the number of closed loops $h$. 
The total number of vertices is $V=\sum_p V_p$. Each propagator gives a power of $g_s$, 
while each interaction vertex with $p$ legs 
gives a power of $g_p/g_s$. The fatgraph will then give a factor
\be
\label{fatfactor}
g_s^{E-V}N^h \prod_p g_p^{V_p}.
\ee
The key point now is to regard the fatgraph as a Riemann surface with holes, in which each
closed loop represents the boundary of a hole. The genus $g$ 
of such a surface is determined by the
elementary topological relation
\be
2g-2 =E-V-h
\ee
therefore we can write (\ref{fatfactor}) as
\be
\label{topfactor}
g_s^{2g-2+h}N^h\prod_p g_p^{V_p}=g_s^{2g-2}t^h \prod_p g_p^{V_p}
\ee
where we have introduced the 't Hooft parameter 
\be
\label{thooftpar}
t=N g_s 
\ee
The fatgraphs with $g=0$ are called {\it planar}, while the
ones with $g>0$ are called {\it nonplanar}. The graph giving the $N^3$ contribution in \figref{ncube} is planar:
it has $E=3$, $V_3=2$ and $h=3$, therefore $g=0$, and it is a sphere with
three holes. The graph in \figref{nsolo} is nonplanar: it has $E=3$, $V_3=2$ and $h=1$, therefore
$g=1$, and represents a torus with one hole (it is easy to see this by drawing the diagram
on the surface of a torus). 

We can now organize the computation of the different quantities in the matrix model 
in terms of fatgraphs. For example, the computation of the free energy is
given in the usual perturbative expansion by connected vacuum bubbles. When the
vacuum bubbles are written in the double line notation, we find that the
perturbative expansion of the free energy is given by
\begin{equation}
F=
\sum_{g=0}^{\infty} \sum_{h=1}^{\infty} F_{g,h} g_s^{2g-2} t^h,
\label{openf}
\end{equation}
where the coefficients $F_{g,h}$ (which depend on the coupling constants of the model 
$g_p$) takes into account the symmetry factors of the different 
fatgraphs. We can now formally define the free energy at genus $g$, $F_g(t)$, 
by keeping $g$ fixed and summing over all closed loops $h$ as in (\ref{thooftre}),
so that the total free energy can be written as
\be
\label{totalf}
F=\sum_{g=0}^{\infty} F_g(t) g_s^{2g-2}.
\ee
This is the {\it genus expansion} of the free energy of the matrix model. 
In (\ref{totalf}) we have written the diagrammatic series as an
expansion in $g_s$ around $g_s=0$, keeping the 't Hooft parameter $t=g_s N$ fixed. 
Equivalently, we can regard it as an
expansion in $1/N$, keeping $t$ fixed, and then the $N$ dependence appears
as $N^{2-2g}$. Therefore, for $t$ fixed and $N$ large, the leading contribution comes
from planar diagrams with $g=0$, which go like ${\cal O}(N^2)$. The nonplanar diagrams give subleading
corrections. Notice that $F_g(t)$, which is the contribution to $F$ to a given order in $g_s$, is given by an
infinite series where we sum over all possible numbers of holes $h$, weighted by $t^h$.

{\bf Example}. One can show that
$$
\langle ({\rm Tr}\, M^3)^2 \rangle_G= g_s^3(12 N^3 + 3 N),
$$   
where the first term corresponds to the two planar diagrams shown in \figref{leading}  
(contributing $3N^3$ and $9 N^3$, respectively), and the
second term corresponds to the nonplanar diagram shown in \figref{nsolo}.
Therefore, in the cubic matrix model the expansion of the
free energy reads, at leading order,
\be
F-F_G={2 \over 3} g_s g_3^2 N^3 + {1\over 6}g_s g_3^2  N + \cdots
\ee

There is an alternative way of writing the matrix model partition function which is very useful.
The original matrix model variable has $N^2$ real parameters, but using the gauge symmetry we can
see that, after modding out by gauge transformations, there are only $N$ parameters left. We can for
example take advantage of our gauge freedom to diagonalize the matrix $M$
\be
\label{gfix}
M \rightarrow U M U^{\dagger} =D,
\ee
with $D={\rm diag}(\lambda_1, \cdots, \lambda_N)$, impose this as a gauge choice, and 
use standard Faddeev-Popov techniques in order to
compute the gauge-fixed integral (see for example \cite{biz}). The gauge fixing (\ref{gfix})
leads to the delta-function constraint
\be
\delta ({}^U M)=\prod_{i<j} \delta^{(2)}({}^U M_{ij})
\ee
where ${}^U M=U M U^{\dagger}$. We then introduce
\be
\label{deltam}
\Delta^{-2}(M)=\int dU \, \delta({}^U M).
\ee
It then follows that the integral of any gauge-invariant function $f(M)$ can be written
as
\be
\int dM \, f(M)=\int dM \, f(M) \Delta^2(M)\int dU \, \delta ({}^U M) = \Omega_N \int \prod_{i=1}^N d\lambda_i
\Delta^2(\lambda) f(\lambda),
\label{fpred}
\ee
where we have used the gauge invariance of $\Delta (M)$, and
\be
\Omega_N =\int dU
\ee
is proportional to the volume of the gauge group $U(N)$, as we will see shortly. We have to
evaluate the the factor $\Delta(\lambda)$, which can be obtained from (\ref{deltam}) 
by choosing $M$ to be diagonal. If 
$$
F(M)=0
$$
is the gauge-fixing condition, the standard Faddeev-Popov formula gives
\be
\Delta^2 (M) = {\rm det} \Biggl( {\delta F({}^U M) \over \delta A}\Biggr)_{F=0}
\label{fpdet}
\ee
where we write $U=e^A$, and
$A$ is a anti-Hermitian matrix. Since
\be
F_{ij}({}^U D)=(U D U^{\dagger})_{ij}= A_{ij} (\lambda_i-\lambda_j) + \cdots.
\ee
(\ref{fpdet}) leads immediately to
\be
\Delta^2(\lambda)= \prod_{i<j} (\lambda_i-\lambda_j)^2,\ee
the square of the Vandermonde determinant. Finally, we fix the factor $\Omega_N$ as follows. The
Gaussian matrix integral can be computed explicitly by using the Haar measure (\ref{haarmeas}), and 
is simply
\be
\int dM \, e^{-{1 \over 2 g_s} {\rm Tr}\, M^2} = (2 \pi g_s)^{N^2/2}.
\ee
On the other hand, by (\ref{fpred}) this should equal
\be
\Omega_N \int \prod_{i=1}^N d\lambda_i \Delta^2(\lambda) e^{-{1 \over 2 g_s} \sum_{i=1}^N \lambda^2_i}.
\ee
The integral over eigenvalues can be evaluated in various ways, using for example the Selberg
function \cite{mehta} or the technique of orthogonal polynomials that 
we describe in the next subsection, and its value is
\be
g_s^{N^2/2} (2 \pi)^{N/2}G_2 (N+2)\ee
where $G_2(z)$ is the Barnes function, defined by
\be
\label{gaussianvalue}
G_2 (z+1)=\Gamma (z) G_2(z), \,\,\,\,\, G_2(1)=1.
\ee
Comparing these results, we find that
\be
\Omega_N= {(2\pi)^{N(N-1)\over 2}\over G_2 (N+2)}.
\ee
Using now (see for example \cite{ovproof}):
\be
\label{volun}
{\rm vol}(U(N))={ (2\pi)^{ {1 \over 2}N(N+1)} \over G_2(N+1)}.
\ee
we see that
\be
\label{gfixint}
{1 \over {\rm vol}(U(N))}\int dM\, f(M)= {1 \over N!}{1 \over (2\pi)^N} \int \prod_{i=1}^N d\lambda_i \, \Delta^2(\lambda)
f(\lambda).
\ee
The factor $N!$ in the r.h.s. of (\ref{gfixint}) has an obvious interpretation: after fixing the
gauge symmetry of the matrix integral by fixing the diagonal gauge, there is still a residual symmetry
given by the Weyl symmetry of $U(N)$, which is the symmetric group $S_N$ acting as permutation of the
eigenvalues. The ``volume'' of this discrete gauge group is just its order, $|S_N|=N!$, and since we
are considering gauged matrix models we have to divide by it as shown in (\ref{gfixint}). As a particular 
case of the above formula, it follows that one can write the partition function (\ref{matrix}) as
\be
\label{parteigen}
Z={1 \over N!} {1 \over (2\pi)^N}\int \prod_{i=1}^N d\lambda_i \, \Delta^2(\lambda)
e^{-{1\over 2 g_s} \sum_{i=1}^N W(\lambda_i)}.
\ee
The partition function of the gauged Gaussian matrix model (\ref{gmm}) is given essentially by the inverse of the 
volume factor. Its free energy to all orders can be computed by using the asymptotic expansion of 
the Barnes function
\ben
\log\, G_2 (N+1)&=& {N^2\over 2} \log \, N -{1\over 12} \log \, N -{3\over 4} N^2 +{1\over 2} N \, \log \, 2\pi + \zeta'(-1)
\nonumber\\
&+& \sum_{g=2}^{\infty} {B_{2g} \over 2g(2g-2)}N^{2-2g},
\een
where $B_{2g}$ are the Bernoulli numbers. 
Therefore, we find the following expression for the total free energy:
\ben
\label{fegmm}
F_G&= &{N^2 \over 2} \Bigl( \log (Ng_s) -{3 \over 2} \Bigr)
-{1 \over 12}\log N + \zeta' (-1) \nonumber \\
&+& \sum_{g=2}^{\infty} {B_{2g} \over
2g (2g-2)} N^{2-2g}.
\een
If we now put $N=t/g_s$, we obtain the following expressions for $F_g(t)$:
\ben
F_0(t)&=&{1\over 2} t^2  \Bigl( \log \, t -{3 \over 2} \Bigr), \nonumber\\
F_1(t)&=&-{1\over 12} \log \, t, \nonumber\\
F_g(t)&=& {B_{2g} \over
2g (2g-2)} t^{2-2g}, \quad g>1.\nonumber
\een
\subsection{Matrix model technology I: saddle-point analysis}

The computation of the functions $F_g(t)$ in closed form seems a difficult task, since 
in perturbation theory they involve summing up an infinite number of fatgraphs (with different numbers of holes $h$). 
However, in the classic paper \cite{brezin} it was shown that, remarkably, $F_0(t)$ can be obtained by 
solving a Riemann-Hilbert problem. In this section we will review this procedure.

Let us consider a general matrix model with action $W(M)$, and let us
write the partition function after reduction to eigenvalues (\ref{parteigen}) 
as follows:
\be
\label{inteff}
Z={1 \over N!} \int \prod_{i=1}^N {d\lambda_i \over 2 \pi}e^{N^2 S_{\rm eff} (\lambda)}
\ee
where the effective action is given by
\be
S_{\rm eff}(\lambda)= -{1 \over tN} \sum_{i=1}^N W(\lambda_i) +
{2 \over N^2}\sum_{i< j}\log |\lambda_i -\lambda_j|.\ee
Notice that, since a sum over $N$ eigenvalues is roughly of order $N$, the effective
action is of order ${\cal O}(1)$. We can now regard $N^2$ as a sort of $\hbar^{-1}$ in such a way
that, as $N\rightarrow \infty$, the
integral (\ref{inteff}) will be dominated by a saddle-point configuration that
extremizes the effective action. Varying $S_{\rm eff} (\lambda)$ w.r.t. the eigenvalue
$\lambda_i$, we obtain the equation
\be
\label{sadd}
{1 \over 2 t} W'(\lambda_i) ={1\over N} \sum_{j \not= i} {1 \over \lambda_i - \lambda_j}, \quad i=1, \cdots, N.
\ee
The eigenvalue
distribution is formally defined for finite $N$ as
\be
\rho (\lambda)={1 \over N} \sum_{i=1}^N \delta (\lambda- \lambda_i),
\label{rhofinite}
\ee
where the $\lambda_i$ solve (\ref{sadd}).
In the large $N$ limit, it is reasonable to expect that this 
distribution becomes a continuous function with compact 
support. We will assume that $\rho(\lambda)$ vanishes outside an interval 
$\cal C$. This is the so-called {\it one-cut solution}.

Qualitatively, what is going on is the following. Assume for simplicity
that $W(x)$, the potential,
has only one minimum $x_*$. We can regard the eigenvalues as
coordinates of a system of $N$ classical particles moving on the real line.
The equation (\ref{sadd}) says that these particles are subject to an effective potential
\be
W_{\rm eff}(\lambda_i)= W(\lambda_i) -{2 t \over N}\sum_{j \not=i} \log |\lambda_i -\lambda_j|
\ee
which involves a logarithmic Coulomb
repulsion between eigenvalues. For small 't Hooft parameter, the potential term dominates over
the Coulomb repulsion, and the particles tend to be in the minimum $x_{*}$ of the
potential $W'(x_{*})=0$. This means that, for $t= 0$, the interval $\cal C$ collapses
to the point $x_*$. As $t$ grows, the Coulomb repulsion will force the eigenvalues to be apart from each
other and to spread out over an interval ${\cal C}$.

We can now write the saddle-point equation in terms of continuum quantities, by using the standard rule
\be
{1\over N}
\sum_{i=1}^N f(\lambda_i)\rightarrow \int_{\cal C} f(\lambda)\rho(\lambda) d\lambda.
\ee
Notice that the distribution of
eigenvalues $\rho (\lambda)$ satisfies the normalization condition
\be
\label{normrho}
\int_{\cal C} \rho (\lambda) d\lambda =1.
\ee
The equation (\ref{sadd}) then becomes
\be
\label{saddrho}
{1 \over 2t}W' (\lambda)= {\rm P} \int {\rho(\lambda') d \lambda' \over \lambda -\lambda'}
\ee
where ${\rm P}$ denotes the principal value of the integral. The above equation
is an integral equation that allows one in principle to compute $\rho(\lambda)$, given
the potential $W(\lambda)$, as a function of the 't Hooft parameter $t$ and the
coupling constants. Once $\rho(\lambda)$ is known, one can easily compute $F_0(t)$: in
the saddle-point approximation, the free energy is given by
\be
\label{freesad}
{1\over N^2}F = S_{\rm eff}(\rho) + {\cal O}(N^{-2}),
\ee
where the effective action in the continuum limit is a functional of $\rho$:
\be
S_{\rm eff}(\rho)=-{1\over t} \int_{\cal C} d\lambda \rho (\lambda) W(\lambda) +
\int_{\cal C \times \cal C} d\lambda d\lambda' \rho(\lambda)\rho(\lambda')\log |\lambda -\lambda'|.
\ee
Therefore, the planar free energy is given by
\be
F_0(t)= t^2 S_{\rm eff}(\rho),
\ee
Since the effective action is evaluated on the distribution of eigenvalues which
solves (\ref{saddrho}), one can simplify the expression to
\be
F_0(t)=-{t\over 2}\int_{\cal C} d\lambda \rho (\lambda) W(\lambda).
\ee
Similarly, averages in the matrix model can be computed in the planar
limit as
\be
{1\over N} \langle {\rm Tr}\, M^{\ell} \rangle =\int_{\cal C}  d\lambda \, \lambda^{\ell} \rho(\lambda).
\label{averho}
\ee
We then see that the planar limit is characterized by a {\it classical} density of states $\rho(\lambda)$, and 
the planar piece of quantum averages can be computed as a moment of this density. The fact that 
the planar approximation to a quantum field theory can be regarded as a classical field 
configuration was pointed out in \cite{master} (see \cite{coleman} for a beautiful exposition). This classical configuration is 
often called the {\it master field}. In the case of matrix models, the master field configuration 
is given by the density of eigenvalues $\rho (\lambda)$, and as we will see 
later it can be encoded in a complex algebraic curve with a deep geometric meaning. 

The density of eigenvalues is obtained as a solution to the saddle-point equation (\ref{saddrho}).
This equation is a singular
integral equation which has been studied in detail in other contexts of
physics (see, for example, \cite{georgia}). The way to solve it
is to introduce an auxiliary function called the {\it resolvent}. The
resolvent is defined as a correlator in the matrix
model:
\be
\omega(p)={ 1\over N} \langle {\rm Tr} {1 \over p -M} \rangle,
\ee
which is in fact a generating functional of the correlation functions (\ref{averho}):
\be
\omega(p) ={1 \over N} \sum_{k=0}^{\infty} \langle {\rm Tr}M^k \rangle p^{-k-1}
\ee
Being a generating functional of connected correlators, it admits an expansion of the form \cite{coleman}:
\be
\omega(p)=\sum_{g=0}^{\infty} g_s^{2 g} \omega_g (p),
\ee
and the genus zero piece can be written in terms of the eigenvalue density as
\be
\label{zeroresint}
\omega_0(p) =\int d \lambda {\rho (\lambda)\over p -\lambda}
\ee
The genus zero resolvent (\ref{zeroresint}) has three important properties. First of all, as a function
of $p$ it is an analytic function
on the whole complex plane
except on the interval $\cal C$, since if $\lambda \in \cal C$ one has a singularity at $\lambda=p$.
Second, due to the normalization property of the eigenvalue distribution (\ref{normrho}), it
has the asymptotic behavior
\be
\label{asymres}
\omega_0(p) \sim {1 \over p}, \qquad p\rightarrow \infty.
\ee
Finally, one can compute the discontinuity of $\omega_0(p)$ as one crosses the
interval $\cal C$. This is just the
residue at $\lambda=p$, and one then finds the key equation
\be
\label{rhow}
\rho(\lambda) =-{1 \over 2 \pi i} \bigl(\omega_0(\lambda+ i\epsilon) -\omega_0 (\lambda-i \epsilon)\bigr).
\ee
Therefore, if the resolvent at genus zero is known, the eigenvalue distribution follows from
(\ref{rhow}), and one can compute the planar free energy. On the other hand, by looking again
at the resolvent as we approach the discontinuity, we see that the r.h.s. of (\ref{saddrho}) is
given by $-(\omega_0(p+ i\epsilon) +\omega_0 (p-i \epsilon))/2$, and we then find the equation
\be
\label{wdisco}
\omega_0(p+ i\epsilon) +\omega_0 (p-i \epsilon)=-{1 \over t} W'(p),
\ee
which determines the resolvent in terms of the potential. In this way we have reduced the 
original problem of computing $F_0(t)$ to the Riemann-Hilbert problem of computing $\omega_0(\lambda)$. 
There is in fact a closed expression for the
resolvent in terms of a contour integral \cite{migdal} which
is very useful. Let $\cal C$ be given by
the interval $b\le \lambda \le a$. Then, one has
\be
\label{solwo}
\omega_0(p) ={1 \over 2t} \oint_{\cal C} {d z \over 2 \pi i} { W'(z) \over p-z} \biggl( { (p-a)(p-b)\over
(z -a) (z -b)}\biggr)^{1 \over 2} .
\ee
This equation is easily proved by converting (\ref{wdisco}) into a discontinuity equation:
\be
{\widehat \omega}_0 (p+i\epsilon) -{\widehat \omega}_0 (p-i\epsilon)=-{1 \over t}
{W'(p) \over {\sqrt {(p-a)(p-b)}}},
\ee
where ${\widehat \omega}_0(p)= \omega_0(p)/{\sqrt {(p-a)(p-b)}}$. This equation determines
$\omega_0(p)$ to be given by (\ref{solwo}) up
to regular terms, but because of the asymptotics (\ref{asymres}), these regular terms are absent.
The asymptotics of $\omega_0(p)$ also gives two more conditions. By taking
$p \rightarrow \infty$, one finds that the r.h.s. of (\ref{solwo}) behaves
like $c + d/p+{\cal O}(1/p^2)$. Requiring the asymptotic behavior (\ref{asymres})
imposes $c=0$ and $d=1$, and this leads to
\begin{eqnarray}
\label{endpo}
\oint_{\cal C} {dz \over 2\pi i} {W'(z) \over {\sqrt {(z-a)(z-b)}}}&=&0, \nonumber\\
\oint_{\cal C}{dz \over 2\pi i} {z W'(z) \over {\sqrt {(z-a)(z-b)}}}&=&2t.
\end{eqnarray}
These equations are enough to determine the endpoints of the cuts, $a$ and $b$, as 
functions of the 't Hooft coupling $t$ and the coupling constants of the model. 

The above expressions are in fact valid for very general potentials 
(we will apply them to logarithmic potentials in section 4), but when $W(z)$ 
is a polynomial, one can find a very convenient expression for the 
resolvent: if we deform the contour in (\ref{solwo}) we pick up a pole at $z=p$, and another 
one at infinity, and we get
\be
\label{wpot}
\omega_0(p)={1 \over 2t} W'(p) -{1 \over 2 t}  {\sqrt {(p-a)(p-b)}}M(p),
\ee
where
\be
M(p)=\oint_0 {dz \over 2 \pi i} {W'(1/z)\over 1-pz} {1 \over{\sqrt {(1-az)(1-bz)}}}.
\ee
Here, the contour is around $z=0$. These formulae, together with the expressions (\ref{endpo}) for the 
endpoints of the cut, completely solve the one-matrix model with one cut in the planar
limit, for polynomial potentials. 

Another way to find the resolvent is to start with (\ref{sadd}),
multiply it by $1/(\lambda_i-p)$, and sum over $i$. One finds, in the limit of
large $N$,
\be
\label{quadw}
(\omega_0(p))^2 - {1\over t} W'(p) \omega_0(p) +{1 \over 4t^2} R(p)=0,
\ee
where
\be
R(p)=4t \int d\lambda \, \rho(\lambda){W'(p) - W'(\lambda) \over p- \lambda}.
\ee
Notice that (\ref{quadw}) is a quadratic equation for $\omega_0(p)$ and has the
solution
\be
\label{quadsol}
\omega_0(p)= {1\over 2t}\Bigl( W'(p) -{\sqrt {(W'(p))^2 - R(p)}} \Bigr),
\ee
which is of course equivalent to (\ref{wpot}).

A useful way to encode the solution to the matrix model is to define
\be
\label{defy}
y(p)= W'(p) -2t\, \omega_0(p).
\ee
Notice that the force on an eigenvalue is given by
\be
\label{force}
f(p)=-W'_{\rm eff}(p)=-{1\over 2}(y(p+i\epsilon)+y(p-i\epsilon)).
\ee
In terms of $y(p)$, the quadratic equation (\ref{quadw}) determining the resolvent can be
written as
\be
\label{hypercurve}
y^2=W'(p)^2 - R(p).
\ee
This is nothing but the equation of a hyperelliptic curve given by a certain deformation 
(measured by $R(p)$)
of the equation $y^2=W'(p)^2$ typical of singularity theory. We will see in the next section that 
this result has a beautiful interpretation in terms of topological string theory 
on certain Calabi-Yau manifolds.

{\bf Example}. {\it The Gaussian matrix model}. 
Let us now apply this technology to the simplest case, the Gaussian model with $W(M) =M^2/2$. 
Let us first look for the position of the endpoints from (\ref{endpo}). Deforming
the contour to infinity and changing $z\rightarrow 1/z$, we find that the first equation
in (\ref{endpo}) becomes
\be
\oint_{0} {d z \over 2 \pi i} {1 \over z^2} {1 \over {\sqrt {(1-az)(1-bz)}}}=0,
\ee
where the contour is now around $z=0$. Therefore $a+b=0$, in accord with the symmetry of the potential.
Taking this into account, the second equation becomes:
\be
\oint_{0} {d z \over 2 \pi i} {1 \over z^3} {1 \over {\sqrt {1-a^2 z^2}}}=2t,\ee
and gives
\be
a=2 {\sqrt t}.
\ee
We see that the interval ${\cal C}=[-a,a]=[-2{\sqrt t} , 2{\sqrt t}  ]$ opens as the 't Hooft
parameter grows up, and as $t\rightarrow 0$ it collapses to the minimum of the potential 
at the origin, as expected. We immediately find from (\ref{wpot})
\be
\omega_0(p)={1\over 2 t} \Bigl(p -{\sqrt { p^2 -4t}}\Bigr),
\ee
and from the discontinuity equation we derive the density of eigenvalues
\be
\rho(\lambda)={1 \over 2 \pi t}{\sqrt { 4t -\lambda^2}}.
\ee
The graph of this function is a semicircle of radius $2{\sqrt t} $, and the above
eigenvalue distribution is the famous Wigner-Dyson semicircle law. Notice also that the equation 
(\ref{hypercurve}) is in this case
\be
y^2=p^2 - 4t.
\ee
This is the equation for a curve of genus zero, which resolves the singularity $y^2=p^2$. We then 
see that the opening of the cut as we turn on the 't Hooft parameter can be interpreted as 
a deformation of a geometric singularity. This will be later interpreted in section 3.5 from the point 
of view of topological string theory on Calabi-Yau manifolds. 

{\bf Exercise}. {\it Resolvent for the cubic matrix model}. Consider the cubic matrix model with potential
$W(M)=M^2/2 + g_3 M^3/3$. Derive an expression for the endpoints of the one-cut solution as a function
of $t$, $g_3$, and find the resolvent and the planar free energy. The solution is worked out in \cite{brezin}.

Although we will not need it in this review, there are well-developed techniques 
to obtain the higher genus $F_g (t) $ as systematic corrections to the saddle-point result $F_0(t)$ \cite{ambjorn, eynard}. 
Interestingly enough, these corrections can be computed in terms of integrals of differentials 
defined on the hyperelliptic curve (\ref{hypercurve}).

We have so far considered the so-called one cut solution to the one-matrix model.
This is not, however, the most general solution, and we now will consider the multicut solution 
in the saddle-point approximation. Recall from our previous discussion that the cut appearing in the one-matrix model
was centered around a minimum of the potential. If the potential has many
minima, one can have a solution with various cuts, centered around the
different minima. The most general
solution has then $s$ cuts (where $s$ is lower or equal than the number 
of minima $n$), and the support of the eigenvalue
distribution is a disjoint union of $s$ intervals
\be
{\cal C}=\cup_{i=1}^s {\cal C}_i,
\ee
where
\be
{\cal C}_i=[x_{2i},x_{2i-1}]
\ee
and $x_{2s}< \cdots < x_1$. The equation (\ref{quadsol}) still gives the solution for the
resolvent, and it is easy to see that the way to have multiple cuts is to require $\omega_0(p)$
to have $2s$ branch points corresponding to the
roots of the polynomial $W'(z)^2-R(z)$. Therefore we have
\be
\omega_0(p)={1 \over 2t} W'(p) -{1 \over 2 t}  {\sqrt {\prod_{k=1}^{2s} (p-x_k)}}M(p),
\ee
which can be solved in a compact way by
\be
\label{solwmulti}
\omega_0(p) ={1 \over 2t} \oint_{\cal C} {d z \over 2 \pi i} { W'(z) \over p-z}
\biggl( \prod_{k=1}^{2s} { p-x_k\over
z-x_k}\biggr)^{1 \over 2}.
\ee
In order to satisfy the asymptotics (\ref{asymres}) the following conditions must hold:
\be
\label{splusone}
\delta_{\ell s}={1\over 2t} \oint_{\cal C} {d z \over 2 \pi i} {z^{\ell} W'(z)
\over \prod_{k=1}^{2s} (z-x_k)^{1\over 2}}, \qquad \ell=0,1, \cdots, s.
\ee
In contrast to the one-cut case, these are only $s+1$ conditions for the $2s$ variables
$x_k$ representing the endpoints of the cut. For $s>1$, there are not enough conditions
to determine the solution of the model, and we need extra input to determine the
positions of the endpoints $x_k$. Usually, the extra condition which is imposed 
is that the different cuts are at equipotential
lines (see for example \cite{bde,akemann}). It is easy to see that in general the effective potential 
is constant on each cut,
\be
W_{\rm eff}(p)=\Gamma_i, \qquad p \in {\cal C}_i,
\ee
but the values of $\Gamma_i$ will be
in general different for the different cuts. This means that there can be eigenvalue
tunneling from one cut to the other. The way to guarantee equilibrium is to choose the
endpoints of the cuts in such a way that $\Gamma_i=\Gamma$ for all $i=1, \cdots, s$.
This gives the $s-1$ conditions:
\be
\label{notun}
W_{\rm eff} (x_{2i+1})=W_{\rm eff}(x_{2i}), \qquad i=1, \cdots, s-1,
\ee
which, together with the $s+1$ conditions (\ref{splusone})
provide $2s$ constraints which allow one to find the positions of the $2s$ endpoints
$x_i$. We can also write the equation (\ref{notun}) as
\be
\int_{x_{2i+1}}^{x_{2i}} dz \, M(z) \prod_{k=1}^{2s}(z-x_k)^{1\over 2} =0.
\ee
In the context of the matrix models describing topological strings, 
the multicut solution is determined by a 
different set of conditions and will be described in section 3.4.
  
\subsection{Matrix model technology II: orthogonal polynomials}

Another useful technique to solve matrix models involves orthogonal polynomials. This technique
was developed in \cite{bessis,biz} (which we follow quite closely), and provides explicit expressions for 
$F_g(t)$ at least for low genus. This technique 
turns out to be particularly useful in the study of the so-called double-scaling limit of matrix models 
\cite{doublescaling}. We will use this technique to study Chern-Simons matrix models, 
in section 4, therefore this subsection can be skipped by the reader who is only interested in the 
conventional matrix models involved in the Dijkgraaf-Vafa approach. 

The starting point of the technique of orthogonal polynomials 
is the eigenvalue representation of the partition function
\be
Z={1 \over N!} \int \prod_{i=1}^N {d\lambda_i \over 2 \pi} \, \Delta^2(\lambda) e^{-{1\over g_s} \sum_{i=1}^N 
W(\lambda_i)},
\ee
where $W(\lambda)$ is an arbitrary potential. If we regard
\be
d\mu = e^{-{1\over g_s} W(\lambda)} {d\lambda \over 2 \pi}
\ee
as a measure
in ${\bf R}$, one can introduce {\it orthogonal polynomials} $p_n(\lambda)$ defined by
\be
\int d\mu \, p_n(\lambda) p_m(\lambda)  = h_n \delta_{nm},\quad n\ge 0,
\label{ortho}
\ee
where $p_n(\lambda)$ are normalized by requiring the behavior $p_n(\lambda)=\lambda^n +\cdots$.
One can now compute $Z$ by noting that
\be
\label{vanderp}
\Delta (\lambda)={\rm det} \, p_{j-1}(\lambda_i).
\ee
By expanding the determinant as
\be
\sum_{\sigma \in S_N} (-1)^{\epsilon(\sigma)} \prod_k p_{\sigma(k)-1}(\lambda_k)
\ee
where the sum is over permutations $\sigma$ of $N$ indices and $\epsilon (\sigma)$ is the signature of the
permutation, we find
\be
\label{parth}
Z =\prod_{i=0}^{N-1} h_i = h_0^N \prod_{i=1}^N r_i^{N-i},
\ee
where we have introduced the coefficients
\be
\label{rcoeff}
r_k= {h_k \over h_{k-1}}, \qquad k\ge 1.
\ee
One of the most important properties of orthogonal polynomials is that they satisfy
recursion relations of the form
\be
\label{recurs}
(\lambda + s_n ) p_n(\lambda) = p_{n+1}(\lambda) + r_n p_{n-1}(\lambda).
\ee
It is easy to see that the coefficients $r_n$ involved in this relation are indeed given by (\ref{rcoeff}). This
follows from the equality
\be
h_{n+1}= \int d\mu \, p_{n+1}(\lambda) \lambda p_n(\lambda),
\ee
together with the use of the recursion relation for $\lambda p_{n+1}(\lambda)$. For even potentials, 
$s_n=0$. 

As an example of this technique, we can consider again the simple case of the 
Gaussian matrix model. The orthogonal polynomials of the 
Gaussian model are well-known: they are essentially the Hermite polynomials $H_n(x)$, which are defined 
by
\be
\label{hermite}
H_n(x)=(-1)^ne^{x^2} {d^n \over dx^n} e^{-x^2}.
\ee
More precisely, one has 
\be
p_n(x)=\Bigl({g_s\over 2}\Bigr)^{n/2} H_n (x/{\sqrt { 2 g_s}}),
\ee
and one can then check that
\be
h_n^G= \Bigl( {g_s \over 2 \pi}\Bigr)^{1\over 2} n! g_s^n, \quad r_n^G=n\, g_s.
\label{gcoefs}
\ee
Using now (\ref{parth}) we can confirm the result (\ref{gaussianvalue}) that we stated before. 

It is clear that a detailed knowledge of the orthogonal polynomials allows the computation of
the partition function of the matrix model. It is also easy to see that the computation of
correlation functions also reduces to an evaluation in terms of the coefficients in the recursion relation. 
To understand this point, it is useful to introduce the {\it orthonormal} polynomials
\be
\label{recurtwo}
{\cal P}_n (\lambda) ={1 \over {\sqrt {h_n}}} p_n(\lambda),
\ee
which satisfy the recursion relation
\be
\lambda {\cal P}_n(\lambda) = -s_n {\cal P}_n(\lambda) + {\sqrt {r_{n+1}}} {\cal P}_{n+1}(\lambda) + {\sqrt {r_n}} {\cal P}_{n-1}(\lambda).
\ee
Let us now consider the normalized vev $\langle {\rm Tr}\, M^{\ell} \rangle$, which in terms of eigenvalues
is given by the integral
\be
\langle {\rm Tr}\, M^{\ell} \rangle={1\over N! Z } \int \prod_{i=1}^N e^{-{1\over g_s}W(\lambda_i)} {d\lambda_i 
\over 2 \pi}
\Delta^2 (\lambda) \Bigl( \sum_{i=1}^N \lambda_i^{\ell} \Bigr).
\ee
By using (\ref{vanderp}) it is easy to see that this equals
\be
\sum_{j=0}^{N-1} \int d\mu \lambda^{\ell} {\cal P}_j^2 (\lambda).
\ee
This integral can be computed in terms of the coefficients in (\ref{recurtwo}).
For example, for $\ell=2$ we find
\be
\langle {\rm Tr}\, M^2 \rangle = \sum_{j=0}^{N-1} (s^2_j + r_{j+1} + r_j),
\ee
where we put $r_0=0$. A convenient way to encode this result is by introducing the
Jacobi matrix
\be
{\cal J}=\begin{pmatrix} 0 & r_1^{1/2} & 0 &0&\cdots\cr
                              r_1^{1/2}& 0 & r_2^{1/2}&0&\cdots\cr
                               0& r_2^{1/2} & 0 & r_3^{1/2}&\cdots\cr
                               \cdots& \cdots&\cdots&\cdots&\cdots \end{pmatrix}
\ee
as well as the diagonal matrix
\be
{\cal S}=\begin{pmatrix}  s_0 & 0 & 0 &0&\cdots\cr
                              0& s_1 & 0&0&\cdots\cr
                               0& 0 & s_2 &0&\cdots\cr
                               \cdots& \cdots&\cdots&\cdots&\cdots\end{pmatrix}.
\ee
It then follows that
\be
\langle {\rm Tr}\, M^{\ell} \rangle={\rm Tr} \, ({\cal J}-{\cal S})^{\ell}.
\label{mnaver}
\ee

The results we have presented so far give the exact answer for the correlators and the partition function,
at all orders in $1/N$. As we have seen, we are particularly interested in computing the functions $F_g(t)$
which are obtained by resumming the perturbative expansion at fixed genus. As shown in \cite{bessis,biz}, one
can in fact use the orthogonal polynomials to provide closed expressions for $F_g(t)$ in the one-cut
case. We will now explain how to do this in some detail.

The object we want to compute is 
\be
{\cal F}=F -F_G=\log \, Z - \log Z_G.
\ee
If we write the usual series ${\cal F}=\sum_{g\ge 0} {\cal F}_g g_s^{2g-2}$, we have
\be
g_s^2 {\cal F}=
{t^2 \over N^2} (\log Z - \log Z_G) = {t^2 \over N} \log {h_0\over h_0^G} + {t^2 
\over N}  \sum_{k=1}^N ( 1-{k\over N})
\log {r_k(N)\over k g_s}.
\label{allf}
\ee
The planar contribution to the free energy ${\cal F}_0(t)$ is obtained from (\ref{allf}) by taking $N \rightarrow \infty$.
In this limit, the variable
$$
\xi={k \over N}
$$
 becomes a continuous variable, $0\le \xi \le 1$, in such a way that
$$
{1\over N} \sum_{k=1}^N f(k/N) \rightarrow \int_0^1 d\xi f(\xi)
$$
as $N$ goes to infinity. Let us assume as well that $r_k(N)$ has the following asymptotic
expansion as $N \rightarrow \infty$:
\be
r_k(N) = \sum_{s=0}^{\infty} N^{-2s} R_{2s}(\xi).
\label{rkexp}
\ee
We then find
\be
\label{orthoplanar}
{\cal F}_0(t)= -{1\over 2} t^2 \, \log \, t + t^2 \int_0^1 d \xi (1 -\xi) \log {R_0 (\xi) \over \xi}.
\ee
This provides a closed expression for the planar free energy in terms of the large $N$ limit of the
recursion coefficients $r_k$.

It is interesting to see how to recover the density of states $\rho(\lambda)$ in the saddle-point approximation
from orthogonal polynomials. Let us first try to evaluate (\ref{mnaver}) in the planar approximation, following
\cite{biz}. A simple argument based on the recursion relations indicates that, at large $N$,
\be
({\cal J}^{\ell})_{nn} \sim {\ell!\over (\ell/2)!^2} r_n^{\ell/2}.
\label{bizasy}
\ee
Using now the integral representation
$$
  {\ell!\over (\ell/2)!^2} = \int_{-1}^1 {dy \over \pi} {(2y)^{\ell}
\over {\sqrt { 1 -y^2}}},
$$
we find
$$
{1\over N} \langle {\rm Tr}\, M^{\ell} \rangle =\int_0^1 d\xi \int_{-1}^1{dy \over \pi} {1 \over {\sqrt { 1 -y^2}}}
(2 y R_0^{1/2} (\xi) -s(\xi))^{\ell},
$$
where we have denoted by $s(\xi)$ the limit as $N \rightarrow \infty$ 
of the recursion coefficients $s_k(N)$ which appear in (\ref{recurs}). 
Since the above average can be also computed by
(\ref{averho}), by comparing we find
$$
\rho(\lambda)= \int_0^1 d\xi \int_{-1}^1{dy \over \pi}{1 \over {\sqrt { 1 -y^2}}}\delta \bigl(\lambda-(2yR_0^{1/2}(\xi) -s(\xi))\bigr),
$$
or, more explicitly,
\be
\rho(\lambda) = \int_0^1 {d \xi \over \pi} { \theta [ 4 R_0(\xi) -(\lambda + s(\xi))^2]\over
{\sqrt { 4 R_0 (\xi)-(\lambda + s(\xi))^2}}}.
\label{rhort}
\ee
Here, $\theta$ is the step function. 
It also follows from this equation that $\rho (\lambda)$ is supported on the interval $[b(t),a(t)]$, 
where 
\be
\label{ortend}
b(t)=-2{\sqrt {R_0(1)}}-s(1), \quad a(t)=2{\sqrt {R_0(1)}} 
- s(1).
\ee

{\bf Example}. In the Gaussian matrix model $R_0(\xi)= t \xi$, and $s(\xi)=0$. We then find that the 
density of eigenvalues is supported in the interval $[-2 {\sqrt t}, 2 {\sqrt t}]$ and it is given by 
$$
\rho (\lambda)= {1\over \pi} \int_0^1 d\xi { \theta [ 4\xi t - \lambda^2] \over {\sqrt { 4\xi t-\lambda^2}}}= {1\over 
2\pi t} {\sqrt { 4t -\lambda^2}}$$
which reproduces of course Wigner's semicircle law.

As shown in \cite{bessis, biz}, orthogonal polynomials can be used as well to obtain the higher genus free energies ${\cal F}_g$.
The key ingredient to do that is simply the Euler-MacLaurin formula, which reads
\be
{1\over N} \sum_{k=1}^N f \bigl( {k\over N} \bigr)= \int_0^1 f(\xi) d\xi + {1\over 2N} [f(1) - f(0)] +
 \sum_{p=1}^{\infty} {1\over N^{2p}} {B_{2p}\over (2p)!}[f^{(2p-1)}(1) - f^{(2p-1)}(0)],
\label{highg}
\ee
and should be regarded as an asymptotic expansion for $N$ large which gives a way to compute systematically $1/N$
corrections. We can then use it to calculate (\ref{allf}) at all orders in $1/N$, where 
\be
f(k/N)=\biggl( 1 - {k \over N} \biggr) \log {N r_k (N) \over k},
\ee
and we use the
fact that $r_k$ has an expansion of the form (\ref{rkexp}). In this way, we find for example that
$$
{\cal F}_1 (t) = t^2 \int_0^1 d\xi (1-\xi) {R_2 (\xi) \over R_0 (\xi)}+ {t^2 \over 12}{d\over d\xi} \Bigl[ (1-\xi) \log
{R_0 (\xi)\over \xi}\Bigr]_0^1,
$$
and so on. We will use this formulation in section 4 to compute ${\cal F}_g(t)$ in the matrix model that describes Chern-Simons
theory on ${\bf S}^3$.

It is clear from the above analysis that matrix models can be solved with the method of
orthogonal polynomials, in the sense that once we know the precise form of the coefficients in the
recursion relation we can compute all quantities in an $1/N$ expansion. Since the recursion relation is only known
exactly in a few cases, we need methods to determine its coefficients for general potentials $W(M)$. In the case of
{\it polynomial} potentials, of the form
$$
W(M) =\sum_{p\ge 0} {g_p\over p} {\rm Tr}\, M^p,
$$
there are well-known techniques to obtain explicit results \cite{biz}, see \cite{df,dfgz} for
reviews. We start by rewriting the recursion relation (\ref{recurs}) as
$$
\lambda \, p_n (\lambda)= \sum_{m=0}^{n+1} B_{nm} p_m,
$$
where $B$ is a matrix. The identities
\ben
 r_n \int d\lambda e^{-{1\over g_s} W(\lambda) } W' (\lambda) p_n(\lambda) p_{n-1}(\lambda)&= & n h_n g_s,\nonumber\\
 \int d\lambda {d \over d\lambda}( p_n  e^{-{1\over g_s}W(\lambda) } p_n) &=& 0\een
lead to the matrix equations
\ben
(W'(B))_{n n-1}&=&n g_s, \nonumber\\
(W'(B))_{n n} &=& 0.
\label{vrecur}
\een
These equations are enough to determine the recursion coefficients. Consider for example
a quartic potential
$$
W(\lambda) = {g_2 \over 2} \lambda^2 + {g_4 \over 4} \lambda^4.
$$
Since this potential is even, it is easy to see that the first equation in (\ref{vrecur}) is automatically
satisfied, while the second equation leads to
$$
r_n \bigl\{ g_2 + g_4 ( r_n + r_{n-1} + r_{n+1}) \bigr\} = n g_s
$$
which at large $N$ reads
$$
R_0 (g_2 + 3 g_4 R_0)= \xi t.
$$
In general, for an even potential of the form
\be
W(\lambda)=\sum_{p\ge 0} {g_{2p+2} \over 2p+2} \lambda^{2p+2}
\ee
one finds
\be
\xi t = \sum_{p\ge 0} g_{2p+2} {2p+1 \choose p} R_0^{p+1}(\xi),
\ee
which determines $R_0$ as a function of $\xi$. The above equation is sometimes called --especially in the 
context of double-scaled matrix models-- the {\it 
string equation}, and by setting $\xi=1$ we find an explicit equation for the endpoints of the 
cut in the density of eigenvalues as a function of the coupling constants and $t$. 

{\bf Exercise}. Verify, using saddle-point techniques, that the string 
equation correctly determines the endpoints of the cut. 
Compute $R_0 (\xi)$ for the quartic and the cubic matrix model,
and use it to obtain ${\cal F}_0(t)$ (for the quartic potential, the 
solution is worked out in detail in \cite{biz}). 

\sectiono{Type B topological strings and matrix models}

\subsection{The topological B model}
 
The topological B model was introduced in \cite{topmatter,tftmirror} 
and can be constructed by twisting the ${\cal N}=2$
superconformal sigma model in two dimensions. There are in fact 
two different twists, called the A and the B twist in \cite{topmatter,tftmirror}, 
and in these lectures we will focus on the second one.
A detailed review of topological
sigma models and topological strings can be found in \cite{Hori}.

The topological B model is a theory of maps from a Riemann surface $\Sigma_g$ to 
a Calabi-Yau manifold $X$ of complex dimension $d$. The Calabi-Yau condition arises in order to 
cancel an anomaly that appears after twisting (see for example Chapter 3 of 
\cite{mmthesis} for a detailed analysis of this issue). Indices for the real tangent bundle 
of $X$ will be denoted by $i=1, \cdots, 2d$, while holomorphic and antiholomorphic
indices will be denoted respectively by $I, {\overline I}
=1, \cdots, d$. The holomorphic tangent bundle will be simply denoted 
by $TX$, while the antiholomorphic tangent bundle will be denoted by ${\overline {TX}}$. 
One of the most important properties of Calabi-Yau manifolds (which can 
actually be taken as their defining feature) is that they have a holomorphic, nonvanishing 
section $\Omega$ of the canonical bundle $K_X=\Omega^{3,0}(X)$. Since the section 
is nowhere vanishing, the canonical line bundle is trivial and $c_1(K_X)=0$. We will always consider examples with 
complex dimension $d=3$. 

The field content of the topological B model is the following. First, since it is 
a nonlinear sigma model, we
have a map $x: \Sigma_g \rightarrow X$, which is a scalar, commuting field. 
Besides the field $x$, we 
have two sets of Grassmann fields $\eta^{\overline I}, \theta^{\overline I} \in x^*({\overline {TX}})$, which are
scalars on $\Sigma_g$, and a Grassmannian one-form on $\Sigma_g$, $\rho_{\alpha}^I$, with
values in $x^*(TX)$. We also have commuting auxiliary fields $F^I, F^{\overline I}$ 
(we will follow here the off-shell formulation of \cite{topmatter,Bmatter}). 
The action for the theory is:
\ben
{\cal L}&=&t \int_{\Sigma_g} d^2z \Big[ G_{I \overline J}\big(\partial_z x^I
\partial_{\zb} x^{\overline J}+ \partial_{\zb} x^I \partial_z x^{\overline J} \big) -
\rho_z^I \big( G_{I \overline J}D_{\zb} \eta^{\overline J} + D_{\zb} \theta_I \big) \nonumber \\
& &
\,\,\,\,\,\,\,\, - \rho_{\zb}^I \big( G_{I \overline J}D_z
\eta^{\overline J}- D_z \theta_I \big)- R^I{}_{J \overline L K}\eta^{\overline L}
\rho_z^{J} \rho_{\zb} ^K \theta_I -G_{I \overline J}F^I F^{\overline J} \Big],
\een
In this action, we have picked local
coordinates $z, \bar z$ on $\Sigma_g$, and $d^2 z$ is the measure $-i dz \wedge d\bar z$. $t$ is a parameter
that plays the role of $1/\hbar$, the field $\theta_I$ is given by $\theta_I=G_{I \overline J}\theta^J$, 
and the covariant derivative $D_{\alpha}$ acts on sections $\psi^i$ of the tangent bundle as
\be
D_{\alpha} \psi^i =\partial_{\alpha} \psi^i + \partial_{\alpha} x^j \Gamma^i_{jk}\psi^k.
\ee
The theory also has a BRST, or topological, charge
$\CQ$ which acts on the fields according to
\be
\begin{array}{cclcccl}
[{\CQ}, x^I]  &=& 0,&
\,\,\,\,\,\,\,\,\,\,\,\,\,  & [\CQ, x^{\overline I}]&=& \eta^{\overline I}, \nonumber\\  
\{ {\CQ},\eta^{\overline I} \}  &=& 0 ,&
\,\,\,\,\,\,\,\,\,\,\,\,\, & \{\CQ,\theta_{I}\} &=&  G_{I \overline J} F^{\overline J},
\nonumber\\ 
\{\CQ,\rho_z^I\}  &=&\partial_z x^I ,& \,\,\,\,\,\,\,\,\,\,\,\,\,  &
 [\CQ,F^I]&=& D_z\rho_{\zb}^I - D_{\zb}\rho_z^I + R^I{}_{J \overline L K}
\eta^{\overline L}
\rho_z^{J} \rho_{\zb} ^K , \nonumber\\
\{\CQ,\rho_{\zb}^I\} &=&\partial_{\zb} x^I,&
\,\,\,\,\,\,\,\,\,\,\,\,\,  &
[\CQ, F^{\overline I}] &=& -\Gamma_{\overline J\overline
K}^{\overline I} \eta^{\overline J} F^{\overline K},
\end{array}
\label{qtrans}
\ee 
The action of $\CQ$ explicitly depends on the 
splitting between holomorphic and antiholomorphic coordinates on $X$, in other 
words, it depends explicitly on the choice of complex structure on $X$. 
It is easy to show that $\CQ^2=0$, and that the action of the model is $\CQ$-exact:
\be
\label{qexact}
 {\cal L}=\{ \CQ, V \}
\ee
where $V$ (sometimes called the gauge fermion) is given by
\be
V=t\int_{\Sigma_g} d^2z \, \big[ G_{I \bar J} \big( \rho_z^I
\partial_{\zb} x^{\bar
 J}+\rho_{\zb}^I \partial_z x^{\bar J} \big) - F^I \theta_I \big].
\ee
Finally, we also have a $U(1)$ ghost number symmetry, in which
$x$, $\eta$, $\theta$ and $\rho$ have
ghost numbers $0$, $1$, $1$, and $-1$, respectively. The Grassmannian charge
$\CQ$ then has ghost number $1$. Notice that, if we interpret $\eta^{\overline I}$ as a basis 
for antiholomorphic differential forms on $X$, the action of ${\cal Q}$ on $x^I$, $x^{\overline I}$ 
may be interpreted as the Dolbeault antiholomorphic differential ${\overline {\partial}}$.

It follows from (\ref{qexact}) that the energy-momentum tensor of this theory 
is given by
\begin{equation}
T_{\alpha \beta}= \{\CQ, b_{\alpha \beta} \},
\label{qex}
\end{equation}
where $b_{\alpha \beta}= \delta V/\delta g^{\alpha \beta}$ and has ghost number $-1$. The fact that
the energy-momentum tensor is $\CQ$-exact means that the theory is
{\it topological}, in the sense that the partition function does not depend on the
background two-dimensional metric. This is easily proved: the partition function
is given by
\be
Z=\int {\cal D} \phi \, e^{-{\cal L}},
\end{equation}
where $\phi$ denotes the set of fields of the theory, and we compute it in the background of a
two-dimensional metric $g_{\alpha \beta}$ on the Riemann surface. Since $T_{\alpha \beta}=
\delta {\cal L}/\delta g^{\alpha \beta}$, we find that
\be
{\delta Z \over \delta g^{\alpha \beta}}=-\langle \{\CQ, b_{\alpha \beta} \} \rangle,
\end{equation}
where the bracket denotes an unnormalized vacuum expectation value. Since $\CQ$
is a symmetry of the
theory, the above vacuum expectation value vanishes, and we find that $Z$ 
is metric-independent, at least formally.

The $\CQ$-exactness of the action itself also has an important consequence:
the same argument that we used above implies that the partition function
of the theory is independent of $t$. Now, since $t$ plays the role of $1/\hbar$, the limit of
$t$ large corresponds to the semiclassical approximation. Since the theory does not
depend on $t$, the
semiclassical approximation is {\it exact}. The classical configurations
for the above action are {\it constant} maps  $x:
\Sigma_g \rightarrow X$. Therefore, it follows that path integrals of the above 
theory reduce to integrals over $X$ \cite{tftmirror}.

What are the operators to consider in this theory? Since the
most interesting aspect of this model is the independence w.r.t. to the two-dimensional metric,
we want to look for operators whose correlation functions satisfy this
condition. It is easy to see that the operators in the cohomology of $\CQ$
do the job: topological invariance requires them to be
$\CQ$-closed, and on the other hand they cannot be $\CQ$-exact, since otherwise their
correlation functions would vanish. One can also check that the $\CQ$-cohomology is
given by operators of the form
\begin{equation}
{\cal O}_{\phi}=\phi_{{\overline I}_1 \cdots {\overline I}_p}^{J_1 \cdots J_q} \eta^{{\overline I}_1} \cdots 
\eta^{{\overline I}_p} \theta_{J_1} \cdots \theta_{J_q},
\label{qops}
\end{equation}
where 
\be
\phi= \phi_{{\overline I}_1 \cdots {\overline I}_p}^{J_1 \cdots J_q} dx^{{\overline I}_1}\wedge  \cdots 
\wedge dx^{{\overline I}_p} {\partial \over \partial x^{J_1}} \wedge \cdots \wedge {\partial \over \partial x^{J_q}}
\ee
is an element of $H_{\overline {\partial}}^p(X, \wedge^q TX)$. Therefore,
the $\CQ$-cohomology is in one-to-one correspondence with the twisted Dolbeault cohomology of the
target manifold $X$. We can then consider correlation functions of the form
\be
\label{correl}
\langle \prod_{a}{\cal O}_{\phi_a}\rangle.
\ee
This correlation function vanishes unless the following selection rule is 
satisfied
\be
\label{selrule}
\sum_{a}p_a=\sum_a q_a=d(1-g), 
\ee
where $g$ is the genus of the Riemann surface. This selection rule comes from a $U(1)_L \times U(1)_R$ anomalous 
global current. Due to the arguments presented above, this correlation function can be computed in the 
semiclassical limit, where the path integral reduces to an integration over the target $X$. The product of operators 
in (\ref{correl}) corresponds to a form in $H_{\overline {\partial}}^d(X, \wedge^d TX)$. To integrate such a form 
over $X$ we 
crucially need the Calabi-Yau condition. This arises as follows. 
In a Calabi-Yau manifold we have an invertible map
\ben
\Omega^{0,p}(\wedge^q TX) &\longrightarrow& \Omega^{d-q,p}(X) \nonumber\\
\phi^{I_1 \cdots I_q}_{{\overline J}_1\cdots {\overline J}_p} &\mapsto& \Omega_{I_1 \cdots I_q I_{q+1}\cdots I_d}
\phi^{I_1 \cdots I_q}_{{\overline J}_1\cdots {\overline J}_p} 
\label{mapomega}
\een
where the $(d,0)$-form $\Omega$ is used to contract the indices. Since $\Omega$ is holomorphic, this descends to 
the ${\overline \partial}$-cohomology. It then follows that an element in $H_{\overline {\partial}}^{d}(X, \wedge^d TX)$ 
maps to an element in $H_{\overline {\partial}}^{d}(X)$. After further multiplication by $\Omega$, one can then integrate 
a $(d,d)$-form over $X$. This is the prescription to compute correlation functions like (\ref{correl}). 
A simple and important example of this procedure is the case of a Calabi-Yau threefold, $d=3$, and 
operators associated to forms in $H_{\overline {\partial}}^1(X, TX)$, or by using (\ref{mapomega}), 
to forms in $H_{\overline {\partial}}^{2,1}(X)$. These operators are important since they correspond to 
infinitesimal deformations of the complex structure of $X$. 
The selection rule (\ref{selrule}) says that we have to integrate 
three of these operators, and the correlation function reads in this case
\be
\label{threecorrel}
\langle {\cal O}_{\phi_1} {\cal O}_{\phi_2} {\cal O}_{\phi_3}\rangle = 
\int_X  (\phi_1)_{{\overline J}_1}^{I_1} (\phi_2)_{{\overline J}_2}^{I_2}(\phi_3)_{{\overline J}_3}^{I_3} 
\, \Omega_{I_1 I_2 I_3} dz^{{\overline J}_1} dz^{{\overline J}_2}dz^{{\overline J}_3} \wedge \Omega.
\ee

It turns out that the full information of the correlators (\ref{threecorrel}) at genus zero can be encoded 
in a single function called the {\it prepotential}. We will quickly review here some of the basic results of 
special geometry and 
the theory of the prepotential for the topological B model, and we refer the reader to \cite{cdlo,Hori} for more details. 
The correlation functions in the B model, like for example (\ref{threecorrel}), depend on a choice of complex 
structure, as we have already emphasized. The different complex structures form a moduli space ${\cal M}$ 
of dimension $h^{2,1}$. A convenient parametrization of ${\cal M}$ 
is the following. Choose first a symplectic basis for $H_3(X)$, denoted by $(A_a, B^a)$, 
with $a=0,1,\cdots, h^{2,1}$, and 
such that $A_a \cap B^b =\delta_a^b$. We then define the {\it periods} of the Calabi-Yau manifold as
\be
z_a =\int_{A_a} \Omega, \qquad {\cal F}^a=\int_{B^a}\Omega, \quad a=0, \cdots, h^{2,1}.
\label{operiods}
\ee
Of course, the symplectic group ${\rm Sp}(2h^{2,1}+2,{\bf R})$ acts on the vector $(z^a, {\cal F}_a)$. A basic 
result of the theory of deformation of complex structures says that 
the $z^a$ are (locally) complex projective coordinates for ${\cal M}$. 
Inhomogeneous coordinates can be introduced in a local patch where one of the 
projective coordinates, say $z_0$, is different from zero, and taking
\be
t_a={z_a \over z_0}, \qquad a=1, \cdots, h^{2,1}.
\ee
The coordinates $z_a$ are called {\it special projective coordinates}, and since they parametrize ${\cal M}$ we 
deduce that the other set of periods must depend on them, {\it i.e.} ${\cal F}^a={\cal F}^a(z)$. 
Using the periods (\ref{operiods}) we can define the {\it prepotential} 
${\cal F}(z)$ by the equation
\be
{\cal F}={1\over 2} \sum_{a=0}^{h^{2,1}} z_a {\cal F}^a.
\label{prepot}
\ee
The prepotential satisfies
\be
{\cal F}^a(z)={\partial {\cal F} \over \partial z_a}
\ee
and turns out to be a homogeneous function of degree two in the $z_a$. Therefore, one can 
rescale it in order to obtain a function of the inhomogeneous coordinates $t_a$:
\be
F_0(t_a)={1\over z_0^2} \, {\cal F}(z_a).
\ee
The fact that the coordinates $z_a$ are projective is related to the freedom in normalizing the three-form 
$\Omega$. In order to obtain expressions in terms of the inhomogeneous coordinates $t_a$, we 
simply have to rescale $\Omega \rightarrow {1\over z_0} \Omega$, and the periods $(z_a, {\cal F}^a)$ 
become
\be
(1, t_a,  2F_0 -\sum_{a=1}^{h^{2,1}}t_a {\partial F_0 \over \partial t_a},{\partial F_0 \over \partial t_a}).
\ee
One of the key results in special geometry 
is that the correlation functions (\ref{threecorrel}) can be computed in terms of 
the prepotential $F_0(t)$. Given a deformation of the complex structure parametrized by $t_a$, the corresponding 
tangent vector $\partial/\partial t_a$ is associated to a differential form of type $(2,1)$. This form 
leads to an operator ${\cal O}_a$, and the three-point functions involving these operators turn out to 
be given by       
\be
\langle {\cal O}_{a} {\cal O}_{b} {\cal O}_{c}\rangle={\partial^3 F_0 \over {\partial t}_a{\partial t}_b 
{\partial t}_c}.
\ee
The prepotential $F_0(t)$ encodes the relevant information about the B model on the sphere, and it 
has an important physical meaning, since it gives the four-dimensional 
supergravity prepotential of type IIB string theory 
compactified on $X$ (and determines the leading part of the vector multiplet effective action). 

In order to obtain interesting quantities at higher genus one has to couple 
the topological B model to two-dimensional gravity, using the fact that the structure of the twisted theory is
very close to that of the bosonic string \cite{dvv,csts,bcov}. In the bosonic string, there is
a nilpotent BRST operator, $\CQ_{\rm BRST}$, and the energy-momentum tensor
turns out to be a $\CQ_{\rm BRST}$-commutator: $T(z)=\{\CQ_{\rm BRST}, b(z)
\}$. In addition, there is a ghost number with anomaly $3 \chi (\Sigma_g)=
6-6g$, in such a way that $\CQ_{\rm BRST}$ and $b(z)$
have ghost number $1$ and $-1$, respectively. This is precisely the
same structure that we found in (\ref{qex}), and
the composite field $b_{\alpha \beta}$ plays the role of an antighost. Therefore, one can
just follow the prescription of coupling to gravity for the bosonic string
and define a genus $g\ge 1$ free energy as follows:
\begin{equation}
F_g= \int_{{\overline M}_{g}} \langle \prod_{k=1}^{6g-6} (b, \mu_k)
\rangle,
\label{fg}
\end{equation}
where
\begin{equation}
(b, \mu_k)=\int_{\Sigma_g} d^2 z (b_{zz}(\mu_k)_{\bar z}^{~z} + b_{\bar z \bar z}
({\overline \mu}_k)_{z}^{~\bar z}),
\end{equation}
and $\mu_k$ are the usual Beltrami differentials.
The vacuum expectation value in (\ref{fg}) refers to the path integral over the
fields of the topological B model, and gives a differential form
on the moduli space of Riemann surfaces of genus $g$, ${\overline M}_g$,
which is then integrated over. The free energies $F_g$ of the B model coupled to gravity for 
$g \ge 1$ are also related to variation of complex structures. A target space description of this 
theory, called Kodaira-Spencer theory of gravity, was found in \cite{bcov}, and 
can be used to determine recursively the $F_g$ in terms 
of special geometry data. 

\subsection{The open type B model and its string field theory description}

The topological B model can be formulated as well for open strings, {\it i.e.}, when 
the worldsheet is an open Riemann surface with boundaries $\Sigma_{g,h}$ \cite{csts,ooy}. In order to construct the 
open string version we need boundary conditions (b.c.) for the fields. It turns out that the appropriate 
b.c. for the 
B model are Dirichlet along holomorphic cycles of $X$, $S$, and Neumann in the remaining directions. 
Moreover, one can add Chan-Paton factors to the model, and this is implemented by considering 
a $U(N)$ holomorphic bundle over the holomorphic cycle $S$. The resulting theory can then be interpreted 
as a topological B model in the presence of $N$ topological D-branes wrapping $S$. 
Since we will be interested in finding a spacetime description of the open topological B model, 
we can consider the case in which the branes fill spacetime (the original case considered in \cite{csts}) 
and deduce the spacetime action for lower dimensional branes by dimensional reduction. In the spacetime 
filling case, when $S=X$, the boundary conditions for the fields are $\theta=0$ along $\partial \Sigma_{g,h}$ and 
that the pullback to $\partial \Sigma_{g,h}$ of $*\rho$ vanishes (where $*$ is the Hodge operator).

The open topological B model can also be coupled to gravity following the same procedure that is 
used in the closed case, and one obtains in this way the open type B topological string propagating along the 
Calabi-Yau manifold $X$. We are now interested in providing a description of this model when the $N$ 
branes are spacetime filling. As shown by Witten in \cite{csts}, the most efficient way to do that is 
to use the cubic string field theory introduced in \cite{sft}.    

In bosonic open string
field theory we consider the worldsheet of the string to be an infinite strip parameterized
by a spatial coordinate $0 \le \sigma \le \pi$ and a time coordinate
$-\infty< \tau <\infty$,
and we pick the flat metric $ds^2 =d\sigma^2 + d\tau^2$. We then
consider maps $x: I \rightarrow X$, with $I=[0, \pi]$ and $X$ the target of the string. The
 string field is a functional of open string configurations $\Psi [x(\sigma)]$, of
ghost number one (the string functional depends as well on the ghost fields, but we do
not indicate this dependence explicitly). In \cite{sft}, Witten defines two operations on the space of string functionals.
The first one is the {\it integration}, which 
is defined formally by folding the string around its midpoint and gluing
the two halves:
\be
\int \Psi =\int {\cal D} x(\sigma) \prod_{0\le \sigma \le \pi/2}
\delta[ x(\sigma) -x(\pi -\sigma)] \Psi [x(\sigma)].
\end{equation}
The integration has ghost number $-3$, which is the ghost number of the vacuum. This
corresponds to the usual fact that in open string theory on the disk one has to
soak up three zero modes. One also defines an associative, noncommutative {\it star product}
$\star$ of string functionals through
the following equation:
\be
 \int \Psi_1 \star \cdots \star \Psi_N =  \int \prod_{i=1}^N {\cal D} x_i (\sigma)
\prod_{i=1}^N \prod_{0\le \sigma \le \pi/2}
\delta[ x_i(\sigma) -x_{i+1}(\pi -\sigma)] \Psi_i [x_i(\sigma)],
\ee
where $x_{N+1}\equiv x_1$. The star product simply glues the
strings together by folding them around their midpoints, and gluing
the first half of one with the second half of the following (see for example
the review \cite{tz} for more details), and it doesn't change the
ghost number.
In terms of these geometric operations, the string
field action is given by
\be
S={1 \over g_s}
\int \biggl( {1 \over 2} \Psi \star Q_{\rm BRST} \Psi + {1 \over 3} \Psi \star \Psi
\star \Psi \biggr)
\label{cubicsft}
\end{equation}
where $g_s$ is the string coupling constant. 
Notice that the integrand has ghost number $3$, while the
integration has ghost number $-3$, so that the action (\ref{cubicsft}) has zero
ghost number. If we
add Chan-Paton factors, the string field is promoted to a $U(N)$ matrix of
string fields, and the integration in (\ref{cubicsft}) includes a trace
${\rm Tr}$. The action (\ref{cubicsft}) has all
the information about the spacetime dynamics of open bosonic strings, with
or without D-branes. In
particular, one can derive the Born-Infeld action describing the dynamics
of D-branes from the above action \cite{taylor}.

We will not need all the technology of string field
theory in order to understand open topological strings. The only piece of
relevant information is the following: the string functional is a function
of the zero mode of the string (which corresponds to the position of the string
midpoint), and of the higher oscillators. If we decouple all
the oscillators, the string functional becomes an ordinary function
of spacetime, the $\star$ product becomes the usual product of functions, and
the integral is the usual integration of functions. The decoupling of the
oscillators is in fact the point-like limit of string theory. As we will see, this
is the relevant limit for topological open type B strings on $X$.

We can now exploit
again the analogy between open topological strings and the open bosonic
string that we used to define the coupling of the topological B
model to gravity ({\it i.e.}, that both have a nilpotent
BRST operator and an energy-momentum tensor that is $\CQ_{\rm
BRST}$-exact). Since both theories have a similar structure,
the spacetime dynamics of open topological type B strings is governed 
as well by (\ref{cubicsft}), where $\CQ_{\rm BRST}$ is given in
this case by the
topological charge defined in (\ref{qtrans}), and where the star product and
the integration operation are as in the bosonic string. The construction of the
cubic string field theory also requires the existence of a ghost number symmetry, which
is also present in the topological sigma model in the form of a $U(1)_R$ symmetry, as
we discussed in 3.1. It is convenient to consider the $U(1)_R$ charge of the
superconformal algebra in the Ramond sector, which is shifted by $-d/2$ with
respect to the assignment presented in 3.1 (here, $d$ is the dimension of the
target). When $d=3$ this corresponds to the normalization used in \cite{sft},
in which the ghost vacuum of the $bc$ system
is assigned the ghost number $-1/2$.

In order to provide the string field theory description of open topological type B strings on
$X$, we have to determine the precise content of the string field, the $\star$ algebra and
the integration of string functionals for this particular model. As in the conventional
string field theory of the bosonic string, we have to consider the Hamiltonian description
of topological open strings. We then take $\Sigma$ to be an infinite strip and
consider maps $x: I \rightarrow X$, with $I=[0, \pi]$.
The Hilbert space is made up out of functionals $\Psi[x(\sigma),\cdots]$, where $x$ is a map
from the interval as we have just described, and the $\cdots$ refer to the Grassmann
fields (which play here the r\^ole of ghost fields). Notice that, since $\rho^I_{z,\zb}$ are canonically conjugate 
to $\eta$, $\theta$, we can choose our functional to depend only on $\eta$, 
$\theta$. It is easy to see that the Hamiltonian has the form
\be
H=\int_0^{\pi} d\sigma \biggl(
t G_{ij} {d x^i \over d \sigma}{d x^j \over d \sigma} + \cdots \biggr).
\end{equation}
We then see that string functionals with $dx^i/d\sigma\not=0$ cannot
contribute: as we saw in the previous subsection, the physics is $t$-independent, therefore we can take
$t\rightarrow \infty$. In this limit the functional gets infinitely massive and decouples from the spectrum, 
unless $dx^i/d\sigma=0$. Therefore, the map $x: I \rightarrow X$ has to be constant and in particular it must be a point
in $X$. A similar analysis holds for the Grassmann fields as
well. Since $\theta=0$ at the boundary, it follows that string functionals are functions of the commuting 
zero modes $x^i$ and $\eta^{\overline I}$, and can be written as
\be
\Psi=A^{(0)}(x) + \sum_{p\ge 1} \eta^{{\overline I}_1} \cdots \eta^{{\overline I}_p} A^{(p)}_{{\overline I}_1 \cdots {\overline I}_p}(x).
\label{stringf}
\end{equation}
These functionals can be interpreted as a sum of $(0,p)$-forms on $X$. If we have
$N$ D-branes wrapping $X$, these forms will be valued in ${\rm End}(E)$ (where 
$E$ is a holomorphic $U(N)$ bundle). The $\CQ$ symmetry acts as on these functionals as the 
Dolbeault operator $\overline \partial$ with values in ${\rm End}(E)$. Notice that a differential form
of degree $p$ will have ghost number $p$.

We are now ready to write the string field action for topological open type B strings
on $X$ with $N$ spacetime filling branes. We have seen that the relevant
string functionals are of the
form (\ref{stringf}). Since in string field theory the string field has ghost number one, 
we must have
\be
\Psi = \eta^{\overline I} A_{\overline I}(x),
\end{equation}
where $A_{\overline I}(x)$ is a $(0,1)$-form taking values in the endomorphisms 
of some holomorphic vector bundle $E$. In other words, the string field is just
the $(0,1)$ piece of a gauge connection on $E$. Since the string field only depends on commuting
and anticommuting zero modes, the star product becomes the
wedge products of forms in $\Omega^{(0,p)}({\rm End}(E))$, and the integration of string functionals becomes
ordinary integration of forms on $X$ wedged by $\Omega$. We then have the following dictionary:
\be
\begin{array}{ccc}
 \Psi \rightarrow A, & \,  & \CQ_{\rm BRST}\rightarrow {\overline \partial}\\
\,  & \,  &\, \\
\star \rightarrow \wedge, & \,  & \int \rightarrow \int_X \Omega \wedge.
\end{array}
\end{equation}
The string field action (\ref{cubicsft}) is then given by
\be
S={1\over 2 g_s} \int_X \Omega \wedge{\rm Tr}\biggl(A \wedge  {\overline \partial} A +{2\over 3}A\wedge A \wedge A \biggr).  
\label{hcs}
\end{equation}
This is the so-called {\it holomorphic Chern-Simons action}. It is a rather peculiar quantum 
field theory in six dimensions, but as we will see, when we consider D-branes of lower dimension, 
we will be able to obtain from (\ref{hcs}) more conventional theories by dimensional reduction.   
  
\subsection{Topological strings and matrix models}

We have seen that the spacetime description of the open B model with 
spacetime filling branes reduces to a six-dimensional 
theory (\ref{hcs}). We will see now that, in some circumstances, this 
theory simplifies drastically and reduces to a matrix model. 

In order to simplify the spacetime description one should 
study simple Calabi-Yau manifolds. The simplest example of a local Calabi-Yau threefold is 
a Riemann surface together
with an appropriate bundle over it. The motivation for considering this kind of 
models is the following.
Consider a Riemann surface $\Sigma_g$ holomorphically embedded
inside a Calabi-Yau threefold $X$, and let
us consider the holomorphic tangent bundle of $X$ restricted to $\Sigma_g$. We then have
\be
TX|_{\Sigma_g}=T\Sigma_g \oplus N_{\Sigma_g}
\ee
where $N_{\Sigma_g}$ is a holomorphic rank two complex vector bundle over
$\Sigma_g$, called the normal bundle of $\Sigma_g$, and the CY condition $c_1(X)=0$ gives
\be
\label{normalconst}
c_1(N_{\Sigma_g})=2g-2.
\ee
The Calabi-Yau $X$ ``near $\Sigma_g$" looks precisely like the total space of
the bundle
\be
\label{lcy}
N\rightarrow \Sigma_g
\ee
where $N$ is regarded here as a bundle over $\Sigma_g$ satisfying (\ref{normalconst}).
The space (\ref{lcy}) is an example of a {\it local} Calabi-Yau threefold, and it is
noncompact.

When $g=0$ and $\Sigma_g=\IP^1$ it is possible to be more precise about the bundle $N$. A theorem
due to Grothendieck says that any holomorphic bundle over $\IP^1$ splits
into a direct sum of line bundles (for a proof, see for example \cite{gh}, pp. 516-7). 
Line bundles over $\IP^1$ are all of the form ${\cal O}(n)$,
where $n \in {\bf Z}$. The bundle ${\cal O}(n)$ can be easily described in terms of two charts
on $\IP^1$: the north pole chart, with coordinates $z, \Phi$ for the base and the fiber, respectively, 
and the south pole chart, with coordinates $z', \Phi'$. The change of coordinates is given by
\be
z'=1/z, \quad \Phi'=z^{-n} \Phi.
\ee
We also have that $c_1({\cal O}(n))=n$. We then find that local Calabi-Yau manifolds that 
are made out of a two-sphere together with a bundle over it are all of the form
\be
\label{pone}
{\cal O}(-a) \oplus {\cal O}(a-2) \rightarrow \IP^1, 
\ee
since the degree of the bundles have to sum up to $-2$ due to (\ref{normalconst}). 

Let us now consider the string field theory of type B open topological strings on 
the Calabi-Yau manifold (\ref{pone}). We will consider a situation where we have 
Dirichlet boundary conditions associated to $\IP^1$, in other words, there are 
$N$ topological D-branes wrapping $\IP^1$. Since the normal directions to the D-brane 
worldvolume are noncompact, the spacetime description can be 
obtained by considering the dimensional reduction of the original string field theory action (\ref{hcs}). 
As usual in D-brane physics, the gauge potential $A$ splits into a gauge potential on the worldvolume 
of the brane and Higgs fields describing the motion along the noncompact, transverse directions. In 
a nontrivial geometric situation like the one here, the Higgs fields are sections of the 
normal bundle. We then get three different fields:
\be
A, \quad \Phi_0, \quad \Phi_1, 
\ee
where $A$ is a $U(N)$ $(0,1)$ gauge potential 
on $\IP^1$, $\Phi_0$ is a section of ${\cal O}(-a)$, and 
$\Phi_1$ is a section of ${\cal O}(a-2)$. Both fields, $\Phi_0$ and $\Phi_1$, take values in the 
adjoint representation of $U(N)$. It is easy to see that the action (\ref{hcs}) becomes
\be
S={1\over g_s} \int_{\IP^1} {\rm Tr}\, \bigl( \Phi_0 {\overline D}_A \Phi_1\bigr),
\ee
where ${\overline D}_A ={\overline \partial} + [A, \cdot]$ is the antiholomorphic covariant derivate. Notice that 
this theory is essentially a gauged $\beta \gamma $ system, since $\Phi_0$, $\Phi_1$ 
are quasiprimary conformal fields of dimensions $a/2$, $1-a/2$, respectively.

We will now consider a more complicated geometry. We start with the Calabi-Yau 
manifold (\ref{pone}) with $a=0$, {\it i.e.}
\be
{\cal O}(0)\oplus {\cal O}(-2) \rightarrow \IP^1.
\ee
In this case, $\Phi_0$ is a scalar field on $\IP^1$, while $\Phi_1$ is 
a $(1,0)$ form (since $K_{\IP^1}={\cal O}(-2)$). If we cover $\IP^1$ with two patches with local coordinates $z, z'$ 
related by $z'=1/z$, the fields in the two different patches, $\Phi_0, \Phi_1$, and $\Phi_0', \Phi_1'$ will be 
related by 
\be
\Phi_0'=\Phi_0, \quad \Phi_1'=z^2 \Phi_1.
\label{phisrel}
\ee
We can regard this geometry as a {\it family} of $\IP^1$s located at $\Phi_1'=0$ (the zero 
section of the nontrivial line bundle ${\cal O}(-2)$) parametrized by $\Phi_0=\Phi_0'=x \in 
{\bf C}$. The idea is to obtain a geometry where we get $n$ isolated $\IP^1$s at fixed positions 
of $x$. To do that, we introduce an arbitrary polynomial of degree $n+1$ on $\Phi_0$,
$W(\Phi_0)$, and we modify the gluing rules above as follows \cite{civ}:
\be
\label{glurules}
z'=1/z, \quad \Phi_0'=\Phi_0, \quad \Phi_1'=z^2 \Phi_1 + W'(\Phi_0)z.
\ee
Before, the $\IP^1$ was in a family parameterized by $\Phi_0 \in {\bf C}$. Now,
we see that there are $n$ isolated $\IP^1$s located at fixed positions of $\Phi_0$
given by $W'(\Phi_0)=0$, since this is the only way to have $\Phi_1=\Phi_1'=0$. 

The geometry obtained by imposing the gluing rules (\ref{glurules}) 
can be interpreted in yet another
way. Call $\Phi_0=x$ and define the coordinates
\be
u=2 \Phi_1', \quad v=2 \Phi_1, \quad y=i(2z'\Phi_1'-W'(x)).
\ee
The last equation in (\ref{glurules}) can now be written as
\be
\label{gensing}
uv +y^2 +W'(x)^2=0.
\ee
This is a singular geometry, since there are singularities along the line $u=v=y=0$ for
every $x_*$ such that $W'(x_*)=0$. For example, if $W'(x)=x$, (\ref{gensing}) becomes, after
writing $u,v \rightarrow u-iv, u+iv$
\be
\label{conifold}
u^2 + v^2 + x^2 + y^2=0.
\ee
This Calabi-Yau manifold is called the {\it conifold}, and it is
singular at the origin. For arbitrary polynomials $W(x)$, the equation (\ref{gensing})
describes more general, singular Calabi-Yau manifolds. Notice that locally,
around the singular points $u=v=y=0$, $x=x_*$, the geometry described by (\ref{gensing}) looks
like a conifold (whenever $W''(x_*)=0$). The manifold described by (\ref{glurules}) is obtained
after blowing up the singularities in (\ref{gensing}), {\it i.e.} we modify the geometry 
by ``inflating" a two-sphere $\IP^1$ at each singularity. This process is called resolution of 
singularities in algebraic geometry, and 
for this reason we will call the manifold specified by (\ref{glurules}) the {\it resolved manifold} $X_{\rm res}$.

We can now consider the dynamics of open type B topological strings on 
$X_{\rm res}$. We will consider a situation in which we have in total $N$ D-branes
in such a way that $N_i$ D-branes are wrapped around the $i$-th $\IP^1$, with
$i=1, \cdots, n$. As before, we have three fields in the adjoint representation of
$U(N)$, $\Phi_0$, $\Phi_1$ and the gauge connection $A$. The action describing the
dynamics of the D-branes turns out to be given by
\be
\label{waction}
S ={1\over g_s} \int_{\IP^1} {\rm Tr}\, \biggl( \Phi_1 {\overline D}_A \Phi_0 + \omega W(\Phi_0)
\biggr)
\ee
where $\omega$ is a K\"ahler form on $\IP^1$ with unit volume. This action was derived in
\cite{kklm,dv}. A quick way to see that the modification of the gluing rules due to adding the
polynomial $W'(\Phi_0)$ leads to the extra term in (\ref{waction}) is to use standard techniques
in CFT \cite{dv}. The fields $\Phi_0, \Phi_1$ are canonically conjugate and on the conformal plane
they satisfy the OPE
\be
\label{ope}
\Phi_0(z) \Phi_1 (w) \sim {g_s \over z-w}.
\ee
Let us now regard the geometry described in (\ref{glurules}) as two disks (or conformal
planes) glued through a cylinder. Since we are in the cylinder, we can absorb the factors of
$z$ in the last equation of (\ref{glurules}). The operator that implements the transformation
of $\Phi$ is
\be
U=\exp {1\over g_s} \oint {\rm Tr}\, W(\Phi_0(z)) \, dz,
\ee
since from (\ref{ope}) it is easy to obtain
\be
\Phi'_1=U \Phi_1 U^{-1}.
\ee
We can also write
\be
U=\exp {1\over g_s} \int_{\IP^1} {\rm Tr}\, W(\Phi_0(z)) \omega
\ee
where $\omega$ is localized to a band around the equator of $\IP^1$ (as we will see immediately,
the details of $\omega$ are unimportant, as long as it integrates to $1$ on the two-sphere).

One easy check of the above action is that the equations of motion lead to the geometric
picture of D-branes wrapping $n$ holomorphic $\IP^1$s in the geometry. The gauge connection is
just a Lagrange multiplier enforcing the condition
\be
[\Phi_0, \Phi_1]=0,
\ee
therefore we can diagonalize $\Phi_0$ and $\Phi_1$ simultaneously. The equation of motion
for $\Phi_0$ is simply
\be
{\overline \partial}\Phi_0=0,
\ee
and since we are on $\IP^1$, we have that $\Phi_0$ is a constant, diagonal matrix.
Finally, the equation of motion for $\Phi_1$ is
\be
{\overline \partial}\Phi_1=W'(\Phi_0)\omega,
\ee
and for nonsingular $\Phi_1$ configurations both sides of the equation must vanish
simultaneously, as we can see by integrating both sides 
of the equation over $\IP^1$. Therefore, $\Phi_1=0$ and the constant eigenvalues of $\Phi_0$
satisfy
\be
W'(\Phi_0)=0
\ee
{\it i.e.} they must be located at the critical points of $W(x)$. In general, we will
have $N_i$ eigenvalues of $\Phi_0$ at the $i$-th critical point, $i=1, \cdots, n$, and this
is precisely the D-brane configuration we are considering.

What happens in the quantum theory? In order to analyze it, we will use the approach developed in \cite{bt} 
for the analysis of two-dimensional gauge theories\footnote{I'm grateful to 
George Thompson for very useful remarks on this derivation.}. First of all, we 
choose the maximally Abelian gauge for
$\Phi_0$, {\it i.e.} we write
\be
\Phi_0=\Phi_0^{\bf k} + \Phi_0^{\bf t},
\ee
where $\Phi_0^{\bf t}$ is the projection on the Cartan subalgebra ${\bf t}$, and $\Phi_0^{\bf k}$
is the projection on the complementary part ${\bf k}$. The maximally Abelian gauge is defined by the condition
\be
\Phi_0^{\bf k}=0
\label{gf}
\ee
which means that the nondiagonal entries of $\Phi_0$ are gauge-fixed to be zero. This 
is in fact the same gauge that we used before to write the matrix model in the eigenvalue basis. 
After fixing the gauge the usual Faddeev-Popov techniques lead to a ghost functional determinant given by
\be
{1\over N!}{\rm Det}_{\bf k}({\rm ad}(\Phi_0^{\bf t}))_{\Omega^0(\IP^1)}
\label{deltaghost}
\ee
where the subscript ${\bf k}$ means that the operator $\Phi_0^{\bf t}$ acts 
on the space ${\bf k}$, and the normalization factor $1/N!$ is the inverse of the order 
of the residual symmetry group, namely the Weyl group which permutes the $N$ entries of 
$\Phi^t_0$. The integrand of (\ref{waction}) reads, after gauge fixing,
\be
{\rm Tr}\, \biggl( \Phi_1^{\bf t} {\overline \partial} \Phi_0^{\bf t}
+ W(\Phi_0^{\bf t})\biggr)
+ 2 \sum_{\alpha} A^{\alpha} \Phi_1^{-\alpha}\alpha(\Phi_0^{\bf t}),
\label{gfaction}
\ee
where $\alpha$ are roots, $E_{\alpha}$ is a basis of $\bf k$, and we have
expanded $\Phi_1^{\bf k} =\sum_{\alpha} \Phi_1^{\alpha} E_{\alpha}$ as well as $A^{\bf k}$. We can now
integrate out the $A^{\alpha}$ to obtain 
\be
{1 \over {\rm Det}_{\bf k}({\rm ad}(\Phi_0^{\bf t}))_{\Omega^{1,0}(\IP^1)} }
\prod_{\alpha>0} \delta(\Phi_1^{\alpha}).
\label{deltatric}
\ee
Here we have used the functional generalization of the standard formula $\delta(a x)=|a|^{-1} \delta (x)$. 
We can now trivially integrate over $\Phi_1^{\bf k}$. The inverse determinant in 
(\ref{deltatric}) combines with (\ref{deltaghost}) to produce 
\be
{{\rm Det}_{\bf k}({\rm ad}(\Phi_0^{\bf t}))_{H^0(\IP^1)}\over 
{\rm Det}_{\bf k}({\rm ad}(\Phi_0^{\bf t}))_{H^{1,0}(\IP^1)}} 
\label{quotient}
\ee
where (as usual) nonzero modes cancel (since they are paired by $\partial$) and one ends with the determinants 
evaluated at the cohomologies. Similarly, integrating out $\Phi_1^{\bf t}$ in (\ref{gfaction}) leads to 
${\overline \partial} \Phi_0^{\bf t}=0$,
therefore $\Phi_0^{\bf t}$ must be constant. The quotient of determinants is easy to evaluate
in this case, and one finds
\be
\biggl[ \prod_{i<j} (\lambda_i-\lambda_j)^2 \biggr]^{h^0(\IP^1)-h^{1,0}(\IP^1)},
\ee
where $\lambda_i$ are the constant eigenvalues of $\Phi_0^{\bf t}$. 
Since $h^0(\IP^1)=1$, $h^{1,0}(\IP^1)=0$, 
we just get the square of the Vandermonde determinant and the partition function reads:
\be
Z={1\over N!}\int \prod_{i=1}^N d\lambda_i \prod_{i<j} (\lambda_i-\lambda_j)^2\,
e^{-{1\over g_s}\sum_{i=1}^N W(\lambda_i)}.
\label{finalz}
\ee
In principle, as explained in 
\cite{bt}, one has to include a sum over nontrivial topological sectors of the Abelian gauge field $A^{\bf t}$ in order 
to implement the gauge fixing (\ref{gf}) correctly. Fortunately, in this case the gauge-fixed action 
does not depend on $A^{\bf t}$, and the inclusion of topological sectors is irrelevant. The 
expression (\ref{finalz}) is (up to a factor $(2\pi)^N$) the gauge-fixed version of the matrix model
\be
Z={1\over {\rm vol}(U(N))} \int {\cal D} \Phi \, e^{-{1\over g_s} {\rm Tr}\, W(\Phi)}
\label{finallz}
\ee
We have then derived a surprising result due to Dijkgraaf and Vafa \cite{dv}:
the string field theory action for open topological B strings on the Calabi-Yau manifold
described by (\ref{glurules}) is a matrix model with potential $W(\Phi)$. 

\subsection{Open string amplitudes and multicut solutions}

The total free energy $F(N_i, g_s)$ of topological B strings on the Calabi-Yau (\ref{glurules}) in the 
background of $N=\sum_i N_i$ branes wrapped around $n$ $\IP^1$'s is of the form (\ref{fmulti}), and 
as we have just seen it is given by the free energy of the matrix model (\ref{finallz}). In particular, 
the coefficients $F_{g, h_1, \cdots, h_n}$ can be computed perturbatively in the matrix model. We 
have to be careful however to specify the classical vacua around which we are doing perturbation theory.      
Remember from the analysis of the matrix model that the
classical solution which describes the brane configuration is characterized by having $N_i$
eigenvalues of the matrix located at the $i$-th critical point of the potential $W(x)$. In the 
saddle-point approximation, this means 
that we have to consider a {\it multicut} solution, with eigenvalues ``condensed'' around {\it all} the 
extrema of the potential. Therefore, in contrast to the multicut solution discussed in 2.2, we have that 
(1) all critical points of $W(x)$ have to be considered, and not only the minima, and (2) the number of 
eigenvalues in each cut is not determined dynamically as in (\ref{notun}), but it is rather 
fixed to be $N_i$ in the $i$-th cut. In other words, the integral of the density of 
eigenvalues $\rho(\lambda)$ along each cut equals a {\it fixed} filling fraction $\nu_i=N_i/N$:
\be
\int_{x_{2i}}^{x_{2i-1}} d\lambda \, \rho (\lambda)=\nu_i,
\label{rhofilling}
\ee
where $N=\sum_{i=1}^n N_i$ is the total number of eigenvalues.
Let us introduce the partial 't Hooft couplings
\be
\label{defsi}
t_i=g_s N_i = t \nu_i .
\ee
Taking into account (\ref{rhow}) and (\ref{defy}), we can write (\ref{rhofilling}) as
\be
\label{detsi}
t_i={1\over 4 \pi i } \oint_{A_i} y(\lambda) d\lambda, \quad i=1, \cdots, n,
\ee
where $A_i$ is the closed cycle of the hyperelliptic curve (\ref{hypercurve}) which 
surrounds the cut ${\cal C}_i$. Assuming for
simplicity that all the $t_i$ are different from zero, and taking 
into account that $\sum_i t_i = t$, we see that (\ref{detsi})
gives $n-1$ independent conditions, where $n$ is the number of critical points of $W(x)$. 
These conditions, together with (\ref{splusone}), determine 
the positions of the endpoints $x_i$ as functions of the $t_i$ and the coupling 
constants in $W(x)$. It is clear that the
solution obtained in this way is not an equilibrium solution of 
the matrix model, since cuts can be centered around local maxima and different cuts
will have different values of the effective potential. This is not surprising, since 
we are not considering the matrix model as a quantum mechanical system {\it per se}, but 
as an effective description of the original brane system. The different choices of filling fractions correspond
to different choices of classical vacua for the brane system. 

A subtle issue concerning the above matrix model is the following. The matrix field $\Phi$ 
in (\ref{finallz}) comes from the B model field $\Phi_0$, which is a holomorphic field. Therefore, the matrix integral 
(\ref{finalz}) should be understood as a contour integral, and in order to define the 
theory a choice of contour should be made. This can be done in perturbation theory, by choosing for example 
a contour that leads to the usual results for Gaussian integration, and therefore at this level the 
matrix model is not different from the usual Hermitian matrix model \cite{dv,twistor}. In some cases, however, regarding 
(\ref{finallz}) as a holomorphic matrix model can be clarifying, see \cite{laza} for an exhaustive 
discussion.  

The above description of the multicut solution 
refers to the saddle-point approximation. 
What is the meaning of the multicut solutions from
the point of view of perturbation theory? To address this issue, let us consider for
simplicity the case of the cubic potential:
\be
{1\over g_s} W(\Phi)={1\over 2g_s}  {\rm Tr}\, \Phi^2 + {1\over 3}{\beta\over g_s} {\rm Tr}\, \Phi^3.
\ee
This potential has two critical points, $a_1=0$ and $a_2=-1/\beta$. The most general multicut solution
will have two cuts. There will $N_1$ eigenvalues sitting at $\Phi=0$, and $N_2$ eigenvalues
sitting at $\Phi=-1/\beta$.
The partition function $Z$ of the matrix model is:
\be
Z={1\over N!}\int\prod_{i=1}^N {d\lambda_i\over 2 \pi} \Delta^2(\lambda)e^{-{1\over 2g_s}
\sum_i \lambda_i^2 -{\beta \over 3g_s} \sum_i \lambda_i^3},
\ee
where $\Delta(\lambda)=\prod_{i<j} (\lambda_i-\lambda_j)$
is the Vandermonde determinant. We can now expand the integrand 
around the vacuum with $\lambda_i=0$ for $i=1,\dots, N_1$ and
$\lambda_i=-{1\over \beta}$ for $i=N_1+1,\dots, N$. Denoting the fluctuations by
$\mu_i$ and $\nu_j$, the Vandermonde determinant becomes
\be
\label{vander}
\Delta^2(\lambda)=\prod_{1\le i_1<i_2\le N_1}\!\!\!\! (\mu_{i_1}-\mu_{i_2})^2
\prod_{1\le j_1<j_2 \le N_2}\!\!\!\!(\nu_{j_1}-\nu_{j_2})^2\,
\prod_{{1\le i\le N_1\atop 1\le j\le N_2}}
\Bigl(\mu_{i}-\nu_j + {1\over \beta}\Bigr)^2.
\ee
We also expand the potential around this vacuum and get
\be
\label{action}
W=\sum_{i=1}^{N_1}\left({1\over 2g_s}\mu_i^2+{\beta\over 3g_s}\mu_i^3\right)
-\sum_{i=1}^{N_2}\left({m\over 2g_s}\nu_i^2-{\beta\over 3g_s}\nu_i^3\right)
+{1 \over 6 \beta^2 g_s}N_2.
\ee
Notice that the propagator of the fluctuations around
$-1/\beta$ has the `wrong' sign, since we are expanding around 
a local maximum. The interaction between the two sets of eigenvalues,
which is given by the last factor in (\ref{vander}),
can be exponentiated and included in the action. This
generates an interaction term between the two eigenvalue bands
\be
\label{effe}
W_{\rm int}=2N_1 N_2\log\beta+2\sum_{k=1}^{\infty}{1\over k}
\beta^ k\sum_{i,j}\sum_{p=0}^k  (-1)^p {k \choose p}
\mu_i^{p} \nu_j^{k-p}.
\ee
By rewriting the partition function in terms of matrices instead of
their eigenvalues, we can
represent this model as an effective two-matrix model,
involving an $N_1 \times N_1$
matrix $\Phi_1$, and an $N_2\times N_2$ matrix $\Phi_2$:
\be
\label{efftwo}
Z={1\over {\rm Vol}(U(N_1)) \times {\rm Vol}(U(N_2))}
\int D\Phi_1 D\Phi_2 e^{-W_1(\Phi_1)-W_2(\Phi_2)
-W(\Phi_1, \Phi_2)},
\ee
where
\ben
\label{terms}
W_1 (\Phi_1)&=&+\tr\Bigl({1\over 2g_s}  \Phi_1^2 + {\beta \over 3 g_s} \Phi_1^3\Bigr), \nonumber\\
W_2 (\Phi_2)&=&-\tr\Bigl({1\over 2 g_s}  \Phi_2^2 - {\beta \over 3 g_s} \Phi_2^3\Bigr), \nonumber\\
W_{\rm int}(\Phi_1,\Phi_2)&=&~2\sum_{k=1}^{\infty}{\beta^k\over k}
\sum_{p=0}^k (-1)^p{k\choose p}{\rm Tr}\,\Phi_1^p~
{\rm Tr}\,\Phi_2^{k-p} \nonumber\\
\qquad\quad &+& N_2 W(a_2)+N_1 W(a_1)-2 N_1 N_2\ln \beta.
\een
Here, $\tr\,\Phi_1^0=N_1$, $\tr\,\Phi_2^0=N_2$, $W(a_1)=0$ and
$W(a_2)=1 /(6 g_s \beta^2)$. Although the kinetic term for $\Phi_2$ has the 
`wrong' sign, we can still make sense of the model in 
perturbation theory by using formal Gaussian integration, and this can in 
fact be justified in the framework of holomorphic matrix models \cite{laza}.
Therefore, the two-cut solution of the cubic matrix model can be 
formally represented in terms of an effective two-matrix model. It is now 
straightforward to compute the free energy 
$F_{\rm pert}=\log\bigl(Z(\beta)/Z(\beta=0)\bigr)$ in perturbation 
theory. It can be expanded as
\be
\label{Fpertdef}
F_{\rm pert}=-N_1 W(a_1)-N_2 W(a_2)-2 N_1 N_2 \ln\beta+
\sum_{h=1}^\infty\sum_{g\ge 0}
(g_s \beta^2)^{2g-2+h} F_{g,h}(N_1,N_2)
\ee
where
$F_{g,h}$ is a homogeneous polynomial in $N_1$ and $N_2$
of degree $h$. One finds, up to fourth order in the coupling constant $\beta$, 
the following result \cite{kmt}:
\ben
& &F_{\rm pert}=-N_1 W(a_1)-N_2 W(a_2)-2 N_1 N_2\ln \beta \nonumber\\
&+& g_s \beta^2 \left[ \Bigl({2\over3}N_1^3-5 N_1^2 N_2+5 N_1 N_2^2
-{2\over3}N_2^3\Bigr)+{1\over 6}(N_1-N_2)\right]\nonumber\\
&+& g^2_s \beta^4 \Bigg[\Bigl({8\over3}N_1^4-{91\over3}N_1^3 N_2+59 N_1^2 N_2^2
-{91\over3}N_1 N_2^3+{8\over3}N_2^4\Bigr)+\Bigl({7\over3}N_1^2-{31\over3}N_1 N_2+{7\over3}N_2^2\Bigr)\Bigg] \nonumber \\
&+& \cdots\nonumber\\
\een
From this explicit perturbative computation one can read off the first 
few coefficients $F_{g,h_1, h_2}$. Of course, this procedure can be generalized, and the $n$-cut solution 
can be represented by an effective $n$ matrix model with interactions among the 
different matrices that come from the expansion of the Vandermonde determinant. 
These interactions can be also incorporated in terms of ghost fields, as explained 
in \cite{dgkv}. This makes possible to compute corrections to the saddle-point 
approximation in perturbation theory. One can also use the multicut solution 
to the loop equations \cite{akemann,kostovcft} with minor modifications to compute the genus one 
correction in closed form \cite{kmt,dst,chekhov}. 

\subsection{Master field and geometric transition}

We have seen that the open topological
string amplitudes on the Calabi-Yau manifold $X_{\rm res}$ 
are computed by a multicut matrix model whose planar 
solution (or, equivalently, its master field configuration) is given by a hyperelliptic curve
\be
\label{hypere}
y^2=W'(x)^2 -R(x).
\ee
Moreover, we also saw in (\ref{detsi}) that the partial 't Hooft couplings can be understood 
as integrals around the $A_i$ cycles of this curve, with $i=1, \cdots, n$. Let us 
now compute the variation of the free energy $F_0(t_i)$ when we vary $t_i$. The variation 
w.r.t. $t_i$ (keeping the $t_j$, $j\not=i$, fixed) can be obtained by computing 
the variation in the free energy as we move one eigenvalue from the cut ${\cal C}_i$ to infinity \cite{dv}. This 
variation is given by (minus) the integral of the force exerted on an eigenvalue, as we move it from the endpoint 
of the cut to infinity. The path from the endpoint of ${\cal C}_i$ to infinity, which does not 
intersect the other cuts ${\cal C}_j$, will be denoted by $B_i$. 
Taking into account (\ref{force}), and the fact that $y(p)$ has no discontinuities outside the cuts ${\cal C}_j$, 
we find  
\be
\label{sg}
{\partial F_0\over \partial t_i}=
\int_{B_i} y(x) dx.
\ee
Usually this integral is divergent, but can be easily regularized by taking $B_i$ to run up to 
a cutoff point $x=\Lambda$, and subtracting the divergent pieces as the cutoff $\Lambda$ goes to 
infinity. For example, for the Gaussian matrix model one has 
\be
\label{bint}
{\partial F_0\over \partial t}= \int_{2 {\sqrt t}}^{\Lambda}dx \, {\sqrt {x^2 -4t}}  = t(\log \, t -1)-2 t \log \,\Lambda 
+ {1\over 2} \Lambda^2 + {\cal O}(1/\Lambda^2).
\ee
Therefore, the regularized integral gives $t(\log \, t -1)$, which is indeed the right result. 
It is now clear that (\ref{detsi}) and (\ref{sg}) look very much like the relations (\ref{operiods}) that define 
the periods (therefore the prepotential) in special geometry. What is the interpretation of the appearance of
special geometry?

Recall that our starting point was a Calabi-Yau geometry obtained as a blowup of
the singularity given in (\ref{gensing}). However, there is another way of smoothing out singularities
in algebraic geometry, which is by {\it deforming} them rather than by resolving then.
For example, the conifold singularity given in (\ref{conifold}) can be smoothed out
by deforming the geometry to
\be
\label{defconi}
x^2 + y^2 + u^2 + v^2=\mu.
\ee
This is the so called {\it deformed conifold}.
Geometrically, turning on $\mu$ corresponds to inflating a three-sphere
in the geometry, since the real section of the conifold is indeed an ${\bf S}^3$.
As $\mu \rightarrow 0$, the three-sphere collapses to zero size, so we can interpret the
singularity as arising from a collapsing three-cycle in the geometry. In the more general
singularity (\ref{gensing}), the generic deformation requires turning on a generic polynomial of
degree $n-1$ $R(x)$, and we get the
Calabi-Yau manifold
\be
\label{defgen}
u^2 + v^2 +y^2+W'(x)^2 =R(x).
\ee
We will call this geometry the {\it deformed manifold} $X_{\rm def}$. 
The deformation by $R(x)$ introduces in fact $n$ three-spheres in the geometry, one
for each singularity (recall that
each of the singular points in (\ref{gensing}) is locally like the conifold). The
noncompact Calabi-Yau manifold (\ref{defgen}) has a holomorphic three-form:
\be
\label{tform}
\Omega ={1\over 2 \pi} {dx dy du \over v}
\ee
The three-spheres created by the deformation can be regarded as
two-spheres fibered over an interval in the complex $x$-plane. To see this, let
us consider for simplicity the case of the deformed conifold (\ref{defconi}), with
$\mu$ real. This geometry contains a three-sphere which is given by 
the restriction of (\ref{defconi}) to real values of the variables. 
If we now consider a fixed, real value of $x$ in the
interval $-{\sqrt \mu}<x<{\sqrt \mu}$,
we get of course a two-sphere of radius ${\sqrt {\mu- x^2}}$.
The sphere collapses at the endpoints of the interval,
$x=\pm {\sqrt \mu}$, and the total geometry of the two-sphere
together with the interval $[-{\sqrt \mu},{\sqrt \mu}]$ is
a three-sphere. In the more general case, the curve
$W'(x)^2- R(x)$ has $n$ cuts with endpoints $x_{2i}, x_{2i-1}$, $i=1, \cdots, n$, and
the $n$ three-spheres are ${\bf S}^2$ fibrations over these cuts.

Let us now consider {\it closed} type B topological strings propagating on $X_{\rm def}$. 
As we saw in 3.1, the genus zero theory is determined by the periods of the three-form
$\Omega$ given in (\ref{tform}). We then choose a symplectic basis of three-cycles
${\widehat A}_i$, ${\widehat B}^j$, with ${\widehat A}_i\cap {\widehat B}^j =\delta_i^j$. Here, 
the ${\widehat A}_i$ cycles are the $n$ three-spheres, and they project to cycles $A_i$ surrounding the cut 
${\cal C}_i=[x_{2i},x_{2i-1}]$ in the $x$-plane. 
The ${\widehat B}_i$ cycles are dual cycles which project in the $x$ plane to the 
$B_i$ paths \cite{civ}. The periods of $\Omega$ are then given by
\be
\label{cyper}
t_i={1\over 4 \pi } \oint_{{\widehat A}_i} \Omega, \quad {\partial F_0\over \partial t_i}
 =\int_{{\widehat B}^i} \Omega.
\ee
It is easy to see that these periods reduce to the periods (\ref{defsi}) and 
(\ref{sg}) on the hyperelliptic curve (\ref{hypere}), respectively.
Let us consider again the case of the deformed conifold (\ref{defconi}), which is simpler since there is
only one three-sphere. Let us compute the $A$-period over this three-sphere, which is an ${\bf S}^2$ fibration 
over the cut $[-{\sqrt \mu},{\sqrt \mu}]$, by first doing the integral
over ${\bf S}^2$, and then doing the integral over the cut. Since 
$v={\sqrt {\mu -x^2 -\rho^2}}$, where $\rho^2= y^2 + u^2$, the integral of $\Omega$ over ${\bf S}^2$
is simply
\be
{1\over 2 \pi} \int_{{\bf S}^2} {dy dz \over {\sqrt {\mu -x^2-\rho^2}}}={\sqrt {\mu -x^2}}.
\ee
Therefore, the $A$-period becomes
\be
t={1\over 2\pi }\int_{-{\sqrt \mu}}^{{\sqrt \mu}} y(x) dx,
\ee
where $y$ is now given by $y^2 + x^2= \mu$. This is nothing but the 
A-period (\ref{defsi}) (up to a redefinition $y\rightarrow -i y$). The general case is very similar, and one
finally obtains that the special geometry (\ref{cyper}) of the deformed Calabi-Yau geometry (\ref{defgen})
is equivalent to the planar solution of the matrix model, given by the hyperelliptic curve (\ref{hypere}) and 
the equations for the partial 't Hooft couplings (\ref{detsi}) and the planar free energy (\ref{sg}).

The physical interpretation of this result is that there is an equivalence
between an {\it open} topological string theory on the manifold $X_{\rm res}$, with $N$ D-branes
wrapping the $n$ spheres obtained by blowup, and a {\it closed} topological string theory
on the manifold $X_{\rm def}$, where the $N$ D-branes have disappeared. Moreover, the
't Hooft couplings $t_i$ in the open string theory become geometric periods in the closed string
theory. Since the open topological strings on $X_{\rm res}$ are described by a matrix model, the fact that
the planar solution reproduces very precisely the deformed geometry is
important evidence for this interpretation. This duality relating an open and a closed
string theory is an example of a {\it geometric}, or {\it large N}, transition. Notice that, as 
a consequence of this duality, the 't Hooft resummation of the matrix model corresponds to a 
closed string theory propagating on $X_{\rm def}$. The master field controlling the planar limit 
(which is encoded in the planar resolvent, or equivalently in the quantity $y(\lambda)$) leads to 
an algebraic equation that describes very precisely the {\it target} of the closed string theory dual. 
The large $N$ transition between these two geometries was proposed in \cite{civ}. The fact that the open 
string side can be described by a matrix model was discovered in \cite{dv}. 

\subsection{Extensions and applications}

The results derived above can be extended to more complicated Calabi-Yau 
backgrounds with branes \cite{dvtwo,dvthree}. For example, one can consider 
ADE type geometries with branes wrapping two-spheres \cite{ckv,fiol}, and the string field theory 
description reduces to the ADE matrix models considered in \cite{kostov}. In the one-matrix model 
described before, the master field is given by a hyperelliptic curve $F(x,y)=0$ which is then 
regarded as the Calabi-Yau manifold 
\be
\label{scy}
uv + F(x,y)=0
\ee
in disguise. In some of the 
examples considered in \cite{dvtwo,dvthree}, however, the master field is no longer described by a hyperelliptic curve, but involves 
a more complicated geometry. This geometry is the Calabi-Yau closed string background that is obtained by geometric transition from the 
open string background with branes. A detailed study of the more 
complicated master field geometries that arise in multimatrix models can be found in \cite{ff}. 

Another consequence of the result of Dijkgraaf and Vafa, together with the 
geometric transition of \cite{civ}, is that the Kodaira-Spencer theory of gravity \cite{bcov} 
on the noncompact Calabi-Yau manifold (\ref{defgen}) is equivalent to the 't Hooft resummation of the 
matrix model with potential $W(x)$. For the simple example of the cubic potential, this was 
explicitly checked at genus one in \cite{kmt}. The formalism developed in \cite{eynard} seems 
to be very appropriate to establish this equivalence in detail.  

As we mentioned in the introduction, the main application of the results of Dijkgraaf and Vafa has been the 
computation of effective superpotentials in supersymmetric gauge theories by using matrix model techniques. This 
is based on the fact \cite{bcov,dvthree} that the resummation $F_{0}(t)$ of the open string 
amplitudes is deeply related to the superpotential of the 
gauge theory which can be obtained from string backgrounds with branes. We refer the reader to \cite{afh,ronne} for an exposition of 
these results.

\sectiono{Type A topological strings, Chern-Simons theory and matrix models}

The conceptual structure of what we have 
seen in the B model is the following: first one shows, by using string field theory, 
that the target space description of open topological B strings reduces to a matrix model 
in certain backgrounds. Then one 
solves the model in the planar limit, and a geometry emerges which is interpreted 
as a closed string dual to the original open string theory. Both geometries are related 
by a large $N$ transition. The first transition of 
this type was discovered in the context of topological A strings by Gopakumar and Vafa \cite{gv}. What we will 
do here is to rederive their result by using the language and technology of matrix models. 
The key ingredient is the fact pointed out in \cite{mm} that the partition function 
of Chern-Simons theory can be written in terms of a somewhat exotic matrix model. 
We will only focus on the matrix model aspects of this correspondence. A detailed 
review of Chern-Simons theory and the geometric transition for the A model can be found in \cite{mmreview}.  

\subsection{Solving the Chern-Simons matrix model}

The Chern-Simons action with gauge group $G$
on a generic three-manifold $M$ is defined by
\begin{equation}
S={k \over 4\pi} \int_M {\rm Tr} \Bigl( A\wedge d A + {2 \over 3} A
\wedge A \wedge A \Bigr)
\label{csact}
\end{equation}
Here, $k$ is the coupling constant, and $A$ is a $G$-gauge connection on the
trivial bundle over $M$. We will consider Chern-Simons
theory with gauge group $G=U(N)$. As noticed in \cite{cs},
since the action (\ref{csact}) does not involve the metric, the resulting
quantum theory is topological, at least formally. In particular,
the partition function
\begin{equation}
Z (M)= \int [{\cal D} A]  e^{iS}
\label{partcs}
\end{equation}
should define a topological invariant of $M$.
A detailed analysis shows
that this is in fact the case, with an extra subtlety related to a choice of framing of the 
three-manifold. 

The partition function of Chern-Simons theory can be computed in a variety
of ways. In \cite{cs} it was shown that in fact the theory is exactly solvable by using nonperturbative methods and the relation to the
Wess-Zumino-Witten (WZW) model. In particular, the partition function 
of the $U(N)$ theory on the three-sphere ${\bf S}^3$ is given by 
\begin{equation}
\label{csun}
Z({\bf S}^3)= {1\over (k+N)^{N/2}}
 \sum_{w \in {\cal W}} \epsilon (w)\exp \Bigl(
-{2 \pi i \over k+N} \rho \cdot w (\rho)\Bigr),
\ee
where the sum over $w$ is a sum over the elements of the
Weyl group ${\cal W}$ of $U(N)$, $\epsilon(w)$ is the signature of $w$, and 
$\rho$ is the Weyl vector of $SU(N)$. By using Weyl's denominator formula,
\be
\sum_{w \in {\cal W}} \epsilon (w) e^{w(\rho)\cdot u} =\prod_{\alpha>0}
2 \sinh {\alpha\cdot u \over 2},
\label{wdf}
\end{equation}
where $\alpha$ are positive roots, one finds
\be
Z({\bf S}^3)= {1 \over (k+N)^{N/2}} \prod_{\alpha>0}
2 \sinh \Bigl({(\alpha \cdot \rho) \over 2}  g_s\Bigr)
\label{css}
\end{equation}
where 
\be
\label{cscoup}
g_s={2 \pi i \over k+N}.
\ee
It was found by Witten that open topological type A strings on $T^*{\bf S}^3$ (which is nothing but the 
deformed conifold geometry (\ref{defconi})) in the presence of $N$ D-branes wrapping ${\bf S}^3$ 
are in fact described by $U(N)$ Chern-Simons theory on ${\bf S}^3$ \cite{csts}. This is the type A model analog to the 
fact that open type B strings on the geometry described by (\ref{glurules}) are 
captured by a matrix model, and in both cases this is shown by using open string field theory. The 
free energy of Chern-Simons theory on ${\bf S}^3$ has an expansion of the form (\ref{freeopen}), 
with $g_s$ given in (\ref{cscoup}), 
and the coefficients $F_{g,h}$, which can be computed by standard perturbation theory, have the interpretation 
of open string amplitudes on $T^*{\bf S}^3$.
  
The analogy between the A story and the B story can be taken even further, since 
it turns out that the partition function of Chern-Simons on ${\bf S}^3$, as well as on many
other three-manifolds, can be represented as a matrix integral \cite{mm}. In the case of
${\bf S}^3$ most of the physical
information in $Z({\bf S}^3)$ can be obtained by other means, but for other three-manifolds like
lens spaces and Seifert spaces, the matrix model representation is crucial in order
to extract the coefficients $F_{g,h}$ \cite{mm}. The Chern-Simons matrix model on ${\bf S}^3$ 
gives however a particularly 
clean way to derive the resummed free energies $F_g(t)$ and the geometry of the 
master field, and we will devote the rest of these lectures to presenting this analysis.

In the case of ${\bf S}^3$ 
the easiest way to derive the matrix model representation of the 
Chern-Simons partition function is through direct computation. Consider the following integral:
\be
\label{uncs}
Z_{CS}={ e^{- { g_s \over 12}N(N^2-1)}\over N!}
\int\prod_{i=1}^N {d \beta_i \over 2\pi} \, e^{-\sum_i \beta^2_i/2 g_s}
\prod_{i<j} \Bigl( 2 \sinh {\beta_i - \beta_j\over 2} \Bigr)^2.
\end{equation}
It can easily be  seen that this reproduces the partition function of $U(N)$ Chern-Simons theory on
${\bf S}^3$, given in (\ref{css}), and the derivation is left as an exercise.

{\bf Exercise}. Use the Weyl formula (\ref{wdf}) to write (\ref{uncs}) as a Gaussian integral,
and show that it reproduces (\ref{csun}).

The measure factor in (\ref{uncs})
\be
\label{qvander}
\prod_{i<j} \Bigl( 2 \sinh {\beta_i - \beta_j\over 2} \Bigr)^2
\ee
is not the standard Vandermonde determinant, although it reduces to it for small
separations among the eigenvalues. In fact, for very small $g_s$, the Gaussian potential
in (\ref{uncs}) will be very narrow, forcing the eigenvalues to be 
close to each other, and one can expand the $\sinh$ in (\ref{qvander}) in power series. At leading
order we find the usual Gaussian matrix model, while the corrections to it can be
evaluated systematically by computing correlators in the Gaussian theory. In this 
way one obtains the perturbative expansion of Chern-Simons theory, see
\cite{mm} for details.

Here we will take a slightly different route in order to analyze the model. First of all, we
want to write the above integral as a standard matrix integral with the usual Vandermonde discriminant.
This can be achieved with the change of variables \cite{tierz}
\be
\exp(\beta_i + t) = \lambda_i,
\ee
where $t=Ng_s$, as usual. It is easy to see that the above integral becomes, up to a factor 
$\exp (-N^3 g_s/2)$,
\be
\label{cslog}
Z_{SW}={1\over N!}
\int\prod_{i=1}^N {d \lambda_i \over 2 \pi} \,\Delta^2(\lambda) \,
\exp \Bigl( -\sum_{i=1}^N (\log \lambda_i)^2/2 g_s \Bigr),
\ee
therefore we are considering the matrix model
\be
Z_{SW}={1\over {\rm vol}(U(N))}\int dM\, e^{-{1 \over 2g_s} {\rm Tr}\, (\log M)^2}.
\label{logsqmm}
\ee
We will call this model the {\it Stieltjes-Wigert matrix model}, hence the subscript in (\ref{cslog}) 
and (\ref{logsqmm}). This is because it can be exactly solved with the so-called 
Stieltjes-Wigert polynomials, as we will explain in a moment. 

Matrix integrals with logarithmic potentials are somewhat exotic, but have appeared
before in connection with the Penner model \cite{penner}, with the $c=1$ string at the 
self-dual radius \cite{dmp,im}, and with the $\IP^1$ model \cite{ehy}. 
We want to analyze now the saddle-point approximation to the 
matrix integral (\ref{uncs}), or equivalently to (\ref{cslog}). Since the model in (\ref{cslog}) has 
the standard Vandermonde, we can use the
techniques of section 2.2. Although the formulae there were obtained for a polynomial potential, some 
of them generalize to arbitrary polynomials. In particular, to obtain the resolvent $\omega_0(p)$ we can 
use the formula (\ref{solwo}) with
\begin{equation}
W'(z)={\log z \over z}.
\end{equation}
\begin{figure}[!ht]
\leavevmode
\begin{center}
\epsfysize=4cm
\epsfbox{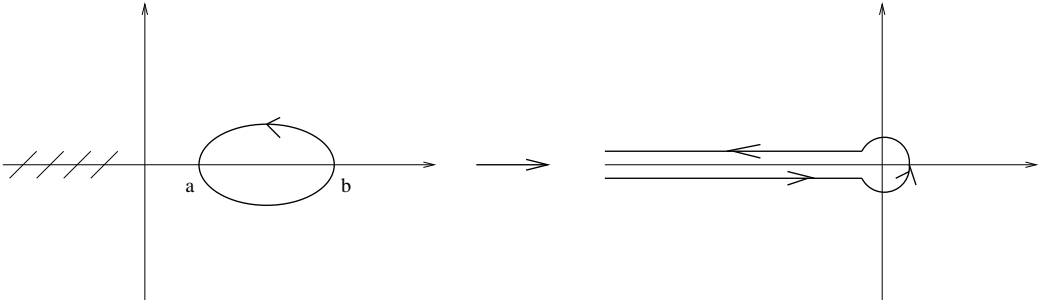}
\end{center}
\caption{This shows the deformation of the contour needed to compute the planar resolvent of the
Chern-Simons matrix integral. We pick a residue at $z=p$, and
we have to encircle the singularity at the origin as well as the branch cut of the logarithm, which
on the left hand side is represented by the dashed lines.}
\label{contour}
\end{figure}
Notice that this potential has a minimum at $z=1$. We then expect a one-cut solution where the
endpoints of the interval $a(t)$, $b(t)$ will satisfy $a(0)=b(0)=1$.
In order to compute the integral (\ref{solwo}) we deform the integration contour. In the case of
polynomial potentials, we picked a residue at $z=p$ and at infinity. Here, since the logarithm
has a branch cut, we cannot push the contour to infinity. Instead, we
deform the contour as indicated in \figref{contour}: we pick the pole at $z=p$, and then we
surround the cut of the logarithm along the negative real axis and the singularity
at $z=0$ with a small circle $C_{\epsilon}$ of radius $\epsilon$. This kind of situation 
is typical of the solution of matrix models with the character expansion \cite{kazakov}.
The resulting integrals are:
\begin{equation}
{1 \over 2t} \Biggl\{ -\int_{-\infty}^{-\epsilon}{dz \over z(z-p) {\sqrt {(z-a)(z-b)}}}
+ \oint_{C_{\epsilon}} {dz \, \log z \over z(z-p) {\sqrt {(z-a)(z-b)}}} \Biggr\}.
\end{equation}
Both are singular as $\epsilon \rightarrow 0$, but singularities
cancel, and after
some computations one finds for the resolvent:
\begin{equation}
\omega_0(p)=-{1 \over 2 tp}
\log \Biggl[ { ({\sqrt a}{\sqrt{ p-b}}-{\sqrt b}{\sqrt {p-a}})^2
\over ({\sqrt {p-a}}-{\sqrt {p-b}})^2 p^2} \Biggr] + {{\sqrt {(p-a)(p-b)}} \over 2tp {\sqrt{ab}}}
\log \Biggl[ {4 ab \over 2 {\sqrt{ab}} + a +b} \Biggr].
\end{equation}
In order to satisfy the asymptotics (\ref{asymres}) the second term must vanish, and the first one
must go like $1/p$. This implies
\begin{eqnarray}
4ab&=& 2 {\sqrt{ab}} + a +b, \nonumber\\
 {\sqrt{a}}+ {\sqrt{b}}&=&2 e^t,
\end{eqnarray}
and from here we obtain the positions of the endpoints of the cut $a, b$ as a function of the
't Hooft parameter:
\begin{eqnarray}\label{endpoints}
a(t)&=&2e^{2t}-e^t +2e^{3 t \over 2}{\sqrt {e^t-1}}, \nonumber\\
b(t)&=&2e^{2t}-e^t -2e^{3 t \over 2}{\sqrt {e^t-1}}.
\end{eqnarray}
Notice that, for $t=0$, $a(0)=b(0)=1$, as expected. The final expression for
the resolvent is then:
\begin{equation}
\omega_0(p)=-{1 \over t p} \log
\Biggl[ {1 + e^{-t} p+ {\sqrt {(1+e^{-t} p)^2-4 p }}  \over 2 p}\Biggr],
\end{equation}
and from here we can easily find the density of eigenvalues
\be
\label{swdensity}
\rho(\lambda) ={1 \over \pi t \lambda} \tan^{-1} \Biggl[ { {\sqrt {4 \lambda -(1+e^{-t} \lambda)^2 }}  \over 1 + e^{-t} \lambda}\Biggr].
\ee
If we now define 
\be
\label{vfun}
u(p)= t(1-p \omega_0(p))+\pi i 
\ee
we see that it solves the equation
\be
e^u +e^v+ e^{v-u+t} +1=0
\label{resconifold}
\ee
where we put $p=e^{t-v}$.
This was found in \cite{akmv} by a similar analysis. The equation (\ref{resconifold}) 
is the analog of (\ref{hypere}) in the case of polynomial matrix models, and can be regarded as 
an algebraic equation describing a noncompact Riemann surface. In fact, (\ref{resconifold}) is 
nothing but the {\it mirror} of the resolved conifold geometry (see for example \cite{horivafa,akv}), and 
$t$ is the K\"ahler parameter of the geometry. 
This is of course in agreement with the result of \cite{gv}, who argued that the 't Hooft resummation 
of Chern-Simons theory leads to a closed string theory propagating on the resolved conifold. As in the B model 
that we analyzed before, the master field of the matrix model encodes the 
information about the target geometry of the closed string description, and provides evidence 
for the geometric transition relating $T^* {\bf S}^3$ and the resolved conifold geometry. 

As we mentioned before, the matrix model (\ref{logsqmm}) can be solved exactly with
a set of orthogonal polynomials called the Stieltjes-Wigert polynomials. The fact that 
the Chern-Simons matrix model is essentially equivalent to the Stieltjes-Wigert matrix model 
was pointed out by Tierz in \cite{tierz}. The Stieltjes-Wigert polynomials are
defined as follows \cite{szego}:
\be
\label{swpol}
p_n (x) = (-1)^n q^{n^2 +{n\over 2}} \sum_{\nu=0}^n \biggl[ {n \atop \nu} \biggr] q^{{\nu (\nu-n)\over 2} - \nu^2}
(-q^{-{1\over 2}} x)^{\nu}
\ee
and satisfy the orthogonality condition (\ref{ortho}) with
\be
\label{swmeas}
d\mu (x) = e^{-{1 \over 2 g_s} (\log \, x)^2} {dx \over 2 \pi}
\ee
and
$$
h_n=  q^{{3 \over 4}n(n+1) +{1 \over 2}} [n]! \Bigl( {g_s \over 2 \pi}\Bigr)^{1\over 2},
$$
where 
\be
\label{qpar}
q=e^{g_s}.
\ee
In the above equations, 
\be
[n]=q^{n\over2} -q^{-{n\over 2}}, \quad  \biggl[ {n \atop m} \biggr]={[n]!\over [m]! [n-m]!}.
\ee
The recursion coefficients appearing in (\ref{recurs}) are in this case
$$
r_n= q^{3n} (q^n-1), \qquad s_n= -q^{{1\over 2} +n} (q^{n+1} + q^n -1).
$$
The Stieltjes-Wigert ensemble can be regarded as a $q$-deformation (in the sense 
of quantum group theory) of the usual Gaussian ensemble. For example, as $g_s \rightarrow 0$ one has 
that $[n]\rightarrow n g_s$, therefore
\be
h_n \rightarrow h_n^G, 
\ee
where $h_n^G$ is given in (\ref{gcoefs}). Also, one can easily check that 
the normalized vev of ${\rm Tr}_R \, M$ 
in this ensemble is given by
\be
\label{swvev}
 \langle {\rm Tr}_R \, M \rangle_{SW}=e^{3t \ell(R)\over 2}q^{\kappa_R\over 2} {\rm dim}_q\, R, 
\ee
where $\ell(R)$ is the number of boxes of $R$, $\kappa_R$ is a quantity defined by
\be
\kappa_R=\ell(R) + \sum_i \lambda_i (\lambda_i-2i)
\ee
in terms of lengths of rows $\lambda_i$ in $R$, and ${\rm dim}_q\, R$ is the quantum dimension 
of the representation $R$
\be
{\rm dim}_q\, R=\prod_{\alpha >0} {[\alpha\cdot (\Lambda + \rho)]\over [\alpha \cdot \rho]}
\ee
where $\Lambda$ is the highest weight associated to $R$. As $g_s \rightarrow 0$, the vev 
(\ref{swvev}) becomes just ${\rm dim}\, R$, the classical dimension of $R$, which is essentially the 
vev in the Gaussian ensemble (\ref{averun}).

Notice that, for this set of orthogonal polynomials, the expansion (\ref{rkexp}) is very simple 
since
\ben
R_0(\xi)&=&e^{4 t \xi}(1 -e^{-t\xi}), \quad R_{2s}(\xi)=0, \,\,\, s>0, \nonumber\\
s(\xi)&=&e^{t \xi} (1-2 e^{t \xi}).
\label{rsw}
\een
As we pointed out in section 2.3, $R_0(\xi)$ and $s(\xi)$ can be used to determine 
the endpoints of the cut in the resolvent through (\ref{ortend}). It is easy to see that (\ref{rsw}) indeed lead 
to (\ref{endpoints}), and that by using (\ref{rhort}) one obtains (\ref{swdensity}). In fact, it is well-known that 
the expression (\ref{swdensity}) is the density of zeroes of the Stieltjes-Wigert polynomials \cite{swdens,cl}. 

We can now use the technology developed in section 2.3 to compute ${\cal F}_g(t)$. Since 
\be
F_{CS}=F_{SW} -{7 \over 12}t^3 + {1\over 12} t,
\ee
the formula (\ref{orthoplanar}) gives
\be
F^{CS}_0(t)= {t^3 \over 12} - {\pi^2 t \over 6} -{\rm Li}_3(e^{-t}) + \zeta(3),
\ee
where the polylogarithm of index $j$ is defined by:
\begin{equation}
{\rm Li}_j (x)= \sum_{n=1}^{\infty} {x^n \over n^j}.
\end{equation}
The above result is in precise agreement with the result in \cite{gv} obtained by 
resumming the perturbative series. With some extra work we can also compute $F^{CS}_g(t)$, for all $g>0$, 
starting from (\ref{highg}). We just have
to compute $f^{(p)} (1) - f^{(p)}(0)$, for $p$ odd, where
$$
f(\xi) = (1-\xi) \phi(\xi,t), \quad \phi(\xi,t)= \log { 1 - e^{-t \xi} \over \xi} + 4t \xi.
$$
It is easy to see that
$$
\phi^{(p)}(\xi,t) = (-1)^{p+1} \Bigl\{ {\rm Li}_{1-p} (e^{-t\xi}) t^p - {(p-1)! \over \xi^p}\Bigr\},
$$
and by using the expansion
$$
{1\over 1- e^{-t}}={1 \over t} + \sum_{k=0}^{\infty} (-1)^{k+1} B_{k+1}  { t ^k \over (k+1)!}
$$
one gets
$$
\phi^{(p)}(0,t)= {(-1)^p B_p \over p} t^p.
$$
Putting everything together, we find for $g>1$
$$
{\cal F}_g (t) = {B_{2g} B_{2g-2} \over 2g (2g-2) (2g-2)!} + {B_{2g} \over 2g (2g-2)!} {\rm Li}_{3-2g}(e^{-t}) -
{B_{2g} \over 2g (2g-2)} t^{2-2g}.
$$
Since the last piece is the free energy at genus $g$ of the Gaussian model, we conclude that the Chern-Simons free energy 
at genus $g$ is given by
\be
F^{CS}_g(t)= {B_{2g} B_{2g-2} \over 2g (2g-2) (2g-2)!} + {B_{2g} \over 2g (2g-2)!} {\rm Li}_{3-2g}(e^{-t})
\label{fgcs}
\ee
which agrees with the resummation of \cite{gv} and also with the genus $g$ closed string amplitude of type A topological 
strings on the resolved conifold (see \cite{mmreview} for more details). 

\subsection{Extensions}

We have seen that the matrix model reformulation of Chern-Simons theory provides an efficient way to obtain the 
master field geometry and to resum the perturbative
expansion. The result (\ref{fgcs}) can be
derived as well from the perturbation series \cite{gv,gvtwo}, but the existence of a matrix model description of 
Chern-Simons theory turns out to be useful in other situations as well. For example, one can easily write
a matrix integral for Chern-Simons theory for other gauge groups \cite{mm}, and the corresponding models
have been analyzed in \cite{hy}. Moreover, the matrix representation of Chern-Simons partition functions 
can be extended to lens spaces and Seifert spaces, and provides a useful way to study perturbative expansions around nontrivial 
flat connections. The matrix models that describe these expansions have been studied in perturbation theory in \cite{mm,akmv} 
and the saddle-point approximation to lens space matrix models has been studied in \cite{hyo}. There are as well multimatrix 
models describing A topological strings on some noncompact Calabi-Yau geometries \cite{akmv} that can be studied by using 
saddle-point techniques \cite{y}, and it is possible 
as well to formulate the Chern-Simons partition function on ${\bf S}^3$ in terms of a unitary model \cite{okuda}. However, all these 
matrix models are usually much harder to analyze than conventional ones, and more work is needed to 
understand their large $N$ properties.

\section*{Acknowledgments}
I would like to thank the organizers of the Les Houches School 
for inviting me to present these lectures in an extraordinary 
environment. I want to thank in particular Volodya Kazakov and 
Paul Wiegmann for the opportunity to lecture in the evenings on Dovjenko, Godard and 
{\it The Matrix} besides my regular lectures on matrix models during the day. I'm also 
grateful to the participants for their enthusiasm, their questions and comments, and the 
fun. Thanks too to Arthur Greenspoon and Niclas Wyllard for a detailed 
reading of the manuscript. Finally, I would like to thank 
Mina Aganagic, Robbert Dijkgraaf, Sergei Gukov, Volodya Kazakov, Albrecht Klemm, Ivan Kostov, Stefan Theisen, 
George Thompson, Miguel Tierz and Cumrun Vafa for educating me about matrix models over the 
last two years.

\end{document}